\documentstyle[epsfig,12pt,aps]{revtex} 
\begin{document}
\newcommand{\p}{\partial}
\newcommand{\ls}{\left(}
\newcommand{\rs}{\right)}
\newcommand{\beq}{\begin{equation}}
\newcommand{\eeq}{\end{equation}}
\newcommand{\beqa}{\begin{eqnarray}}
\newcommand{\eeqa}{\end{eqnarray}}
\newcommand{\bdm}{\begin{displaymath}}
\newcommand{\edm}{\end{displaymath}}

%\renewcommand{\baselinestretch}{1.5}
%\twocolumn[\hsize\textwidth\columnwidth\hsize
%           \csname @twocolumnfalse\endcsname]
%\draft
%%%%%%%%%%%%%%%%%%%%%%%%%%%%%%%%%%%%%%%%%%%%%%%%%%%%%%%%%%%%%%%%%%%
\title{Consequences of kinetic non-equilibrium
for the nuclear equation-of-state in heavy ion collision
}
\author{
C. Fuchs$^{1}$ and T. Gaitanos$^{2}$}
\address{
$^{1}$Institut f\"ur Theoretische Physik der Universit\"at T\"ubingen, 
Auf der Morgenstelle 14, D-72076 T\"ubingen, Germany\\
$^{2}$ Sektion Physik der Universit\"at M\"unchen, 
Am Coulombwall 1, D-85748 Garching, Germany\\
}
\maketitle  
%%%%%%%%%%%%%%%%%%%%%%%%%%%%%%%%%%%%%%%%%%%%%%%%%%%%%%%%%%%%%%%%%%%
\begin{abstract}
Highly compressed nuclear matter created in relativistic heavy 
collisions is to large extent governed by local non-equilibrium.
As an idealized scenario colliding nuclear matter 
configurations are studied within both, relativistic mean field theory 
and using more realistic in-medium interactions based on the Dirac-Brueckner 
T-matrix. The equation of state in anisotropic matter is thereby governed 
by two competing effects: The enlarged phase space volume in 
colliding matter tends 
to soften the internal potential energy of the subsystems whereas the 
relative motion of the two currents leads to a strong additional repulsion 
in the system. An effective EOS constructed for anisotropic momentum 
configurations shows a significant net softening compared to ground 
state nuclear matter. This effect is found to be to large extend 
independent on the particular choice of the nuclear interaction. 
A critical discussion of standard transport approaches 
with respect to the considered non-equilibrium effects is given.
\end{abstract}
%%%%%%%%%%%%%%%%%%%%%%%%%%%%%%%%%%%%%%%%%%%%%%%%%%%%%%%%%%%%%%%%%%%
\section{Introduction}
%%%%%%%%%%%%%%%%%%%%%%%%%%%%%%%%%%%%%%%%%%%%%%%%%%%%%%%%%%%%%%%%%%%
One major goal of relativistic heavy ion physics is 
to explore the behavior of the nuclear equation-of-state (EOS) far 
away from saturation, i.e. at high densities and non-zero 
temperature. Over the last three decades a large variety of observables 
has been investigated both, from the experimental and theoretical 
side motivated by the search for the nuclear EOS. 
The collective particle flow is thereby intimately connected to the 
dynamics during the compressed high 
density phase of such reactions \cite{stoecker86,ritter97}. 
E.g., the elliptic flow which develops in the early 
compression phase is thought to be a suitable observable 
to extract information on the EOS \cite{eflow}. But also the 
production of strange particles is a good probe to study dense 
matter \cite{AiKo85}. Recent precision measurements of the $K^+$ production 
at SIS energies strongly support the scenario of a soft 
EOS \cite{sturm00,fuchs00}. 

However, the difficulty to extract information on the EOS 
$\epsilon (\varrho)$ from 
heavy ion reactions lies in the fact that the system is out 
of equilibrium. At intermediate energies the relaxation time needed 
to equilibrate coincides more or less with the high density phase of 
the reaction. Hence, non-equilibrium effects are present 
over the compression phase where one essentially intends to 
study the EOS at supra-normal densities. 
Experimental evidence for incomplete equilibration 
even in central collisions has recently been reported in \cite{rami99}. 
In the early stages of the reaction, the phase space distribution 
in the participant zone is similar to two fluids of 
counter-streaming matter. The local momentum distributions of such 
colliding nuclear matter, called CNM in the following, are given by 
two Fermi ellipsoids, i.e. two boosted Fermi spheres separated 
by a relative velocity \cite{sehn96}. 
In the course of the reaction the mid-rapidity region 
is more and more populated due to binary collisions and the system 
is heated up. The originally cold and sharp momentum 
ellipsoids become diffuse and merge together. The energy density 
depends in the course of the reaction therefore on both, the {\it local} 
relative velocity and the {\it local} temperature:
$$
\epsilon (\varrho,v_{\rm rel},T=0)\longrightarrow
\epsilon (\varrho,v_{\rm rel},T)\longrightarrow
\epsilon (\varrho,v_{\rm rel}=0,T)~~.
$$
In order to learn something about the ground state EOS one 
has to seperate the compression energy from the contributions 
of the relative motion and the thermal energy
$$
\epsilon (\varrho,v_{\rm rel},T) = \epsilon_{\rm comp} + 
{\cal E}_{\rm rel} + {\cal E}_{\rm therm}
$$
This is generally a difficult task since the various contributions are 
not independent from each other. Moreover, the conversion of 
kinetic energy into compression and thermal energy goes along 
with the production of entropy. While the total energy is conserved, 
the free energy tends to minimize. This temporal evolution 
of a heavy ion reaction is described by microscopic 
transport models like BUU \cite{bkc88} or QMD \cite{Ai91}. 
Such simulations \cite{essler97,lang,puri} show also that 
the local phase space in the overlapping overlapping zone of two 
interpenetrating nuclei can to large extent be idealized by 
colliding nuclear matter. CNM provides a smooth transition 
to a final, at least in the central cell, 
fully equilibrated spherical momentum configuration. 

The aim of the present work is to investigate in more detail 
how the compression energy, related to the {\it ground state} EOS, 
can be extracted in kinetic non-equilibrium situations 
and which consequences arise if one tries to do this 
by the use of dynamical transport simulations. 
In the limit of vanishing temperature the compression 
energy is just the free energy of the system 
$\epsilon_{\rm comp} = \epsilon (\varrho,v_{\rm rel},T=0) - {\cal E}_{\rm rel}$. 
To obtain an effective EOS in CNM one has to 
subtract the energy of the relative motion.  
Such a treatment neglects temperature effects 
but it allows to study non-equilibrium features for simplified 
configurations. These reflect in first order the phase space evolution 
in the initial and the high density phase of the reaction where 
termperatures are still moderate. Therefore 
the infinite counter-streaming two-fluid 
scenario has been a subject of a variety of theoretical 
investigations 
\cite{sehn96,gmat2,wbc92,neise90,ivanov91,elsenhans92,neisethesis,thesis}.

In transport calculations the nuclear mean field is usually 
based on phenomenological parameterizations 
\cite{bkc88,Ai91,welke88,dani00}. Such parameterizations allow 
different extrapolations to high densities, summarized 
by referring to a 'hard' or a 'soft' equation-of-state \cite{stoecker86},  
which can be tested in heavy ion collisions. A more microscopic approach 
is to start from free NN interactions, determine the correlated two-body 
interaction in nuclear matter and use this interaction in 
heavy ion reactions. Doing so, in each time step 
in the course of the reaction the mean field, respectively 
the EOS, would have to be 
derived from an integration of the two-body interaction over 
the actual momentum distribution. This determines then both, the 
density and the momentum dependence of the mean field 
\cite{sehn96,btm90,gmat1,lca}. In phenomenological approaches, even 
when momentum dependent interactions 
are included, part of the arising non-equilibrium 
features is missing. Also with phenomenological two-body interactions 
the complete single particle potential should consistently be 
derived for the actual momentum configuration. 
In standard transport calculation this is, however, not done: 
the single particle potential contains a purely density 
dependent part which, as a density functional, has been determined 
for equilibrium matter. The momentum dependent part, on the other hand, 
accounts in lowest order for anisotropy and temperature 
effects of the phase space distribution. However, the density 
dependent part is that one which is most directly related to 
the ground state EOS. 

To obtain a model independent picture the present 
investigations are based on different choices of the nuclear 
forces. We use the framework of relativistic mean field theory, 
i.e. the Walecka model \cite{sw86,bog82} and 
the microscopic relativistic Dirac-Brueckner (DB) 
approach \cite{thm87,boelting99}. The mean field approach 
has the advantage that most integrals can 
be evaluated analytically which makes the discussion of 
typical features of CNM more transparent. The DB model, 
on the other side, is more realistic. It turned out to be 
quite successful in the description of 
nuclear matter bulk properties. DB forces have been extensively tested at 
SIS energies below 1 A.GeV, and 
a general agreement with corresponding flow data has been observed 
\cite{dani00,lca,fopi95}. Other studies of CNM have been performed in a 
non-relativistic framework using either Skyrme interactions 
\cite{neise90,puri} or Brueckner 
G-matrix forces \cite{gmat2,elsenhans92}. Relativistically 
colliding matter has been mainly treated within the Walecka model 
\cite{sehn96,wbc92,ivanov91,neisethesis}. In \cite{sehn96} a first 
attempt was made to apply relativistic 
DB forces. In some cases also finite temperature effects have been studied 
\cite{thesis,neisethesis}. In  \cite{puri} 
$T$ dependent G-matrix forces were used in transport simulations 
but the influence of the explicit $T$ dependence of the mean field 
on the reaction dynamics was found to be moderate.

The paper is organized as follows: In Sec.2 we discuss the general 
structure of the relativistic energy-momentum tensor in colliding 
nuclear matter within both, the DB and the mean field (MF) 
approach. In Sec.3 we derive an effective EOS for the CNM 
configuration. In Sec.4 implications for heavy ion reactions are 
discussed, in particular with respect to transport 
calculations. In Sec.5 we give some short remarks concerning the 
treatment in non-relativistic approaches and finally conclude in 
Sec.6.

%%%%%%%%%%%%%%%%%%%%%%%%%%%%%%%%%%%%%%%%%%%%%%%%%%%%%%%%%%%%%%%%%%%
\section{Energy-momentum tensor in colliding nuclear matter}
%%%%%%%%%%%%%%%%%%%%%%%%%%%%%%%%%%%%%%%%%%%%%%%%%%%%%%%%%%%%%%%%%%%
The phase space distribution of colliding nuclear matter (CNM) 
\beq
\Theta_{12} = \Theta_{1} +  \Theta_{2} - \Theta_{1}\cdot \Theta_{2}
\label{conf}
\eeq
is composed by the momentum distributions of two counter-streaming matters 
$ \Theta_{i} =  \Theta (\mu^*(k_{F_{i}}) - k_{\nu}^* u_{i}^\nu )$, i.e. 
two boosted Fermi ellipsoids. $ \Theta$ is the step function, $k_{F_{i}}$ are the Fermi 
momenta and $ u_{i}^\nu = (\gamma_i, \gamma_i {\bf u}_i)$ 
are the streaming velocities of the two 
subsystem currents. The last term in eq. (\ref{conf}) ensures that the 
Pauli principle is fulfilled for small velocities where 
the two ellipsoids might overlap. The total baryon density has thereby 
to be restored. Details of this procedure which can be performed by 
a covariant construction can be found in \cite{sehn96}.  
%%%%%%%%%%%%%%%%%%%%%%%%%%%%%%%%%%%%%%%%%%%%%%%%%%%%%%%%%%%%%%%%%%%
\subsection{DB approximation}
%%%%%%%%%%%%%%%%%%%%%%%%%%%%%%%%%%%%%%%%%%%%%%%%%%%%%%%%%%%%%%%%%%%
The interaction of nucleons with the surrounding medium is expressed in terms of 
a self energy which dresses the particles, respectively  the in-medium propagator. 
Analogous to spin-isospin saturated nuclear matter at rest, the self energy 
in colliding matter can be decomposed into 
scalar and vector components expressed covariantly by 
\begin{eqnarray}
\Sigma (k) &=& \Sigma_ {S}(k) - \gamma_\mu \Sigma^{\mu}(k)
\\
\Sigma_{\mu}(k)  & = & \Sigma_{0}(k)  u_{\mu 12} + 
 \Sigma_{v}(k) \Delta^{\mu\nu}_{12} k_\nu
\quad .
\label{S2}
\end{eqnarray}
Here, the $\Sigma_{i}$ with $i=S,0,v$ are scalar functions. 

The four--velocity 
$u^{\mu}_{ 12}$ is the streaming velocity of the combined system derived from 
the total current $j^{\mu}_{ 12}=\varrho_{12} u^{\mu}_{12}$. Since  $\varrho_{12}$ 
is the invariant baryon density in the c.m. frame of the two currents,  
$ \varrho_{12} = \sqrt{ j_{12}^2} = < 1 >_{12} |_{c.m.}$, it is natural to work 
in that frame, i.e. the frame where the sum of the spatial components 
of the two counter-streaming currents vanish ${\bf j}_{12} =0$. With 
the help of the projector perpendicular to $u^{\mu}_{12}$ 
\beq
\Delta^{\mu\nu}_{12} = g^{\mu\nu} - u^{\mu}_{12} u^{\nu}_{12}
\label{proj1}
\eeq
eqs. (\ref{S2}) are manifestly covariant. Inversely, the self-energy 
components are obtained from covariant projections 
\beqa
 \Sigma_{S} &=& \frac{1}{4} {\rm tr}[  \Sigma ] 
\\
 \Sigma_{0} &=& \frac{-1}{4} {\rm tr}\left[ u^{\mu}_{12} \gamma_\mu \Sigma  \right] 
\label{sigma_0_proj2}
\\
\Sigma_{v} &=& \frac{-1}{4\ls  \Delta^{\mu\nu}_{12} k_\mu k_\nu \rs } 
{\rm tr} \left[\Delta^{\mu\nu}_{12} k_\mu \gamma_\nu \Sigma  \right] 
\quad .
\label{sigma_v_proj2}
\eeqa
One should note that in this form the projections are identical to 
nuclear matter at rest \cite{thm87} except for the fact that all 
quantities are determined in the colliding system. Hence the effective 
mass and the kinetic momenta are given by 
\begin{eqnarray}
M^{\ast} & = & M +\Sigma_{S}(k)
\label{M0}  \\
k_{\mu}^\ast & = & k_\mu + \Sigma_{\mu}(k) \\
E^\ast & = & \sqrt{{\bf k}^{\ast 2} + M^{\ast 2} (k)}
\quad .
\label{M1}
\end{eqnarray}
The self-energy has now to be derived from some model, e.g. from an 
effective two-body interaction like the in-medium T-matrix. 
In this general case the self-energy components 
(\ref{S2}-\ref{M1}) are momentum dependent. We also consider the 
mean field approach of the Walecka model \cite{sw86} where 
the self-energies are not explicitely momentum dependent. 
This has the advantage 
that all integrals can be solved analytically. In 
the DB approach it is useful to introduce an additional 
configuration averaged and momentum independent effective mass. 
In the c.m. frame (${\bf j}_{12}=0$) one can thus get rid of the 
$\Sigma_v$ contribution by rescaling the fields in the same way 
as it is done in nuclear matter \cite{thm87,boelting99}. The spatial 
component of the self-energy (\ref{sigma_v_proj2}) has in the c.m. frame 
the form ${\bf \Sigma} =\Sigma_v (k) {\bf k}$. In NM $\Sigma_v$ has been 
found to be small and nearly constant as a function of momentum. In mean 
field theory this contribution vanishes identically. 
Here it can be absorbed into a rescaled effective mass 
\beq
{\tilde M}^*=\frac{M+\overline{\Sigma}_{S}}{1+\overline{\Sigma}_{v}}
\label{refspec}
\eeq  
and the rescaled momentum 
${\tilde {\bf k}}^{*}= {\bf k}^{*}/(1+\overline{\Sigma}_{v})={\bf k}|_{c.m.}$. 
The configuration-averaged scalar and vector self-energy components are 
defined as 
\beqa
\overline{\Sigma}_{S} &=&  
< \Sigma_{S}(k)  {\tilde M}^* / {\tilde E}^{\ast} >_{12} / \varrho_{S_{12}}
\label{scalar}\\
\overline{\Sigma}_{v} &=&  
< \Sigma_{v}(k) / {\tilde E}^{\ast} >_{12} / < 1/ {\tilde E}^{\ast} >_{12}~~.
\eeqa
Here 
\beq
<X >_{12}  = \frac{\kappa}{(2\pi)^3} 
\int d^3 {\bf k} X(k) \Theta_{12} (k;\chi) 
\label{confint}
\eeq
denotes the summation over all occupied states and $ {\tilde E}^{\ast}$ is 
defined as in (\ref{M1}), however, with the rescaled quantities 
${\tilde M}^*$ and ${\tilde {\bf k}}^{*}$. In spin-isospin saturated nuclear 
matter the phase space occupancy factor in (\ref{confint}) 
equals $\kappa =4$. The set of 
parameters which determines the colliding configuration is 
denoted by $\chi = \{ k_{F_{1}},k_{F_{2}}, u_1, u_2, {\tilde M}^* \}$. 
Thus the effective mass $ {\tilde M}^* $,  
the scalar density $\varrho_{S_{12}} = <{\tilde M}^* /{\tilde E}^* >_{12}$ and 
the configuration (\ref{conf}) itself are coupled by non-linear equations. 
The Dirac part of the in-medium propagator has the same 
form as in nuclear matter 
\cite{sw86,horowitz87}
\beq
G_{12} (k)=[\gamma_\mu {\tilde k}^{*\mu}+\tilde{M}^*] 2\pi i \delta (\tilde{k}^{*2}
- \tilde{M}^{*2}) 2\Theta(\tilde{k}^*_0) \Theta_{12}(k;\chi)~~.
\label{diracprop}
\eeq
It should be noticed that the usage of an averaged effective 
mass in the propagator 
is a standard procedure of the DB approach \cite{thm87,boelting99} 
(reference spectrum approximation). It is based  on the observation 
that $M^*$ is only weakly varying with momentum inside the Fermi sea 
and it simplifies the solution of the Bethe-Salpeter equation considerably. 
To account for the full momentum dependence of the effective mass 
in the two-body propagator of the BS-equation is presently beyond 
the scope of DB calculations. 
In CNM this approximation is more severe due to large relative 
momenta which can occur between the two currents. However, 
since the effective mass shows in DB calculations only a rather 
moderate momentum dependence at large momenta above the Fermi surface 
\cite{boelting99} this approximation is 
still justified. The self-energy components follow now from covariant 
projections of the T-matrix elements. 
For on-shell scattering of positive energy states the 
T-matrix can generally be decomposed into five Lorentz invariant 
amplitudes (scalar, vector, tensor, axial vector and pseudo-scalar or 
pseudo-vector, respectively) \cite{horowitz87,thm87} where direct (Hartree) 
and exchange (Fock) amplitudes are related by a Fierz 
transformation \cite{tjon85b}. If the amplitudes are already 
anti-symmetrized the 'direct' representation for the T-matrix is 
sufficient \cite{boelting99} 
\beqa
\Sigma (k)&=& -i \int {{d^4 q}\over{(2\pi)^4}}
{\rm tr} \left[ G_{12} (q) ~T(k,q;\chi)\right]
\nonumber \\
&=& \frac{\kappa}{(2\pi)^3} \int \frac{d^3{\bf q}}{\tilde{E}^*({\bf q})}
\left[\tilde{M}^*  T_{\rm S}(k,q;\chi) 
+\not{\tilde q}^*  T_{\rm V}(k,q;\chi) \right] \Theta_{12} (q;\chi)~~.
\label{self1}
\eeqa
The scalar $T_{\rm S}$ and the vector $T_{\rm V}$ amplitude are 
Lorentz invariant functions which depend (besides the parametric 
dependence on the configuration expressed by $\chi$) on the c.m. 
momentum, the c.m. relative momentum and the c.m. scattering 
angle between $k$ and $q$ in a Lorentz invariant manner which 
can be converted to a dependence on the Mandelstam variables 
$s^*,t^*,u^*$. 

The energy momentum tensor in CNM is given by 
\beqa
T^{\mu\nu} &=& -i \int {{d^4 k}\over{(2\pi)^4}}
{\rm tr} \left[ \gamma ^\mu G_{12} (k) \right] k^\nu 
- \frac{i^2}{2} g^{\mu\nu} \int {{d^4 k}\over{(2\pi)^4}}  
 {{d^4 q}\over{(2\pi)^4}} {\rm tr} \left[ G_{12} (k)~ 
{\rm tr} \left[  G_{12} (q) T(k,q;\chi) \right]\right] 
\nonumber\\
&=&  \frac{\kappa}{(2\pi)^3}
           \int \frac{d^3{\bf k}}{\tilde{E}^*({\bf k})} 
            {\tilde k}^{*\mu} k^\nu \Theta_{12}(k;\chi)
\nonumber\\
&-& \frac{g^{\mu\nu}}{2} \frac{\kappa^2}{(2\pi)^6}
           \int \frac{d^3{\bf k}}{\tilde{E}^*({\bf k})}
                \frac{d^3{\bf q}}{\tilde{E}^*({\bf q})}
\left[  {\tilde M}^{*2} T_S (k,q;\chi) + {\tilde k}^{*}_\mu {\tilde q}^{*\mu}
T_V (k,q;\chi) \right] \Theta_{12}(k;\chi) \Theta_{12}(q;\chi)~.
\label{et1}
\eeqa
With eqs. (\ref{M0}-\ref{M1}) we express $T^{\mu\nu}$ in terms of the fields 
\beq
        T^{\mu\nu} = 
       < {\tilde k}^{\ast \mu} k^{\ast\nu} / {\tilde E}^{\ast} >_{12} 
        - V^{\mu\nu}
        - \frac{1}{2} g^{\mu\nu} \left\{ \overline{\Sigma}_{S}\,
        \varrho_{S_{12}}
        - V^{\lambda}_{\lambda}   \right\}~.
\label{tfull}
\eeq    
The scalar contribution (\ref{scalar}) and the terms 
containing the vector field are given by  
\beq
V^{\mu\nu} = <  {\tilde k}^{\ast\mu} \Sigma^{\nu}(k)  / {\tilde E}^{\ast} >_{12}
\quad .
\label{vector}
\eeq
The total scalar density and the total baryon current are given by 
\beqa
\varrho_{S_{12}} &=& -i \int {{d^4 k}\over{(2\pi)^4}}
{\rm tr} \left[ G_{12} (k) \right] = <{\tilde M}^* /{\tilde E}^* >_{12} \\
j_{\mu\, 12}  &=& -i \int {{d^4 k}\over{(2\pi)^4}}
{\rm tr} \left[ \gamma_\mu G_{12} (k) \right] = <{\tilde k}^{*}_\mu /{\tilde E}^* >_{12}
\label{dens1}
\eeqa
%%%%%%%%%%%%%%%%%%%%%%%%%%%%%%%%%%%%%%%%%%%%%%%%%%%%%%%%%%%%%%%%%%%
\subsection{Mean field approximation}
%%%%%%%%%%%%%%%%%%%%%%%%%%%%%%%%%%%%%%%%%%%%%%%%%%%%%%%%%%%%%%%%%%%
In order to examine the structure of the energy-momentum tensor 
(\ref{et1}) it is instructive to consider first the mean field approximation. 
To do so we use the Walecka model (QHD-I) \cite{sw86}. In  QHD-I the 
T-matrix amplitudes are replaced by corresponding coupling constants 
for a scalar $\sigma$ and a vector $\omega$ meson field, 
$T_S \longmapsto \Gamma_S = g_{\sigma}^2/m_{\sigma}^2$ and 
$T_V \longmapsto \Gamma_V = g_{\omega}^2/m_{\omega}^2$, and thus 
the double integrals in (\ref{et1}) decouple. One could also use 
more sophisticated model such as 
density dependent hadron field theory \cite{fule95} 
where $\Gamma_{S,V}(\varrho)$ 
are density dependent coupling functions. Such a density 
dependence can e.g. be taken from effective 
field theories \cite{FST} or again from the DB T-matrix  \cite{fule95}. In the 
latter case this provides the mean field approximation to the DB problem 
outlined above. In the following we consider the simpler case of constant 
couplings $\Gamma_{S,V}$ but the considerations can easily be generalized. 
Since the self-energies are now momentum independent 
one has ${\tilde k}^* = k^*$ and ${\tilde M}^* = M^*$. 
The energy-momentum tensor reads
\beq
        T^{\mu\nu} = 
       < k^{\ast \mu} k^{\ast\nu} / E^{\ast} >_{12} 
        - \Gamma_V j^{\mu}_{12}  j^{\nu}_{12} 
        - \frac{1}{2} g^{\mu\nu} \left\{ \Gamma_S \varrho_{S_{12}}\varrho_{S_{12}}
        - \Gamma_V j^{\lambda}_{12}j_{\lambda\,12} \right\}~~.
\label{tm1}
\eeq  
In nuclear matter one obtains the kinetic energy and pressure densities as 
\beqa
\epsilon_{\rm kin} &=& < E^{\ast} ({\bf k} )>  
 = \frac{3}{4}E_F \varrho (k_F) + \frac{1}{4}M^* \varrho_{S} \\
p_{\rm kin} &=& \frac{1}{3} < {\bf k}^{\ast 2} / E^{\ast} >  
 = \frac{1}{4} E_F \varrho (k_F) - \frac{1}{4}M^* \varrho_{S} ~~~.
\eeqa
The corresponding expressions in CNM are similar and transparent as 
long as the momentum distributions of the two currents do not overlap. 
Otherwise the contributions arising from the Pauli correction in (\ref{conf}) 
complicate the expressions considerably. Thus we first discuss 
the case where the relative velocity of the 
two currents is large enough to seperate the ellipsoids, 
i.e. $\Theta_1 (k,u_1)\Theta_2 (k,u_2) =0$. 
The integrals for the kinetic energy density can  
be solved analytically and yield \cite{sehn96}
\beq
< k^{\ast \mu} k^{\ast\nu} / E^{\ast} >_{12} = 
g^{\mu\nu} \left\{ \frac{1}{4}M^*\varrho_{S_{12}} 
+ \sum_{i=1,2} \frac{3}{4} E_{F_i} \varrho_0 (k_{F_i}) 
\right\} -\sum_{i=1,2} \Delta_{i}^{\mu\nu} E_{F_i} \varrho_0 (k_{F_i}) 
\label{tm2}
\eeq  
Here $\varrho_0 (k_{F_i})$ and $E_{F_i}=\sqrt{k_{F_i}^2 +M^{*2}}$ are the 
rest densities and the corresponding Fermi energies of the subsystems 
and $ \Delta_{i}^{\mu\nu}$ is the projector (\ref{proj1}) on the 
subsystem velocities. To further simplify (\ref{tm2}) we consider 
symmetric configurations ($k_{F_{1,2}}=k_F,u_{1,2}=\pm u $). 
The total (c.m.) current is then given by 
$j_{\mu\,12} =j_{\mu\,1}+j_{\mu\,2} = (2\gamma(u) \varrho_0 (k_F), {\bf 0})$. 
Analogously the total scalar density reads
\beqa
\varrho_{S_{12}} =\varrho_{S_{1}} + \varrho_{S_{2}} = 2\varrho_{S}(M^*) 
= \frac{2\kappa}{4\pi^2} M^* \left[ k_F E_F -M^{*2} {\rm ln}
\left( \frac{k_F + E_F }{M^*}\right)\right]
\label{rs1}
\eeqa
with $M^* = M-\Gamma_S \varrho_{S_{12}}$. 
Energy density $\epsilon = T^{00}$, transverse pressure 
$p_\perp = T^{11}=T^{22}$ and longitudinal pressure 
$p_\parallel = T^{33}$ are given by
\beqa
T^{00} &=& 2\left[ \frac{3}{4} E_F \varrho_0 (k_F) 
       + \frac{1}{4} M^* \,\varrho_{S}(M^*) 
       + E_F \varrho_0 (k_F)\, (\gamma^2 -1)  \right]
\nonumber \\
       &+& \frac{1}{2}\left\{ \Gamma_V \,(2\gamma\varrho_0 (k_F))^2 
       + \Gamma_S\, (2\varrho_{S}(M^*) )^2 \right\}
\label{tm3}\\
p_\perp &=& 2\left[ \frac{1}{4} E_F \varrho_0 (k_F) 
       - \frac{1}{4} M^* \,\varrho_{S}(M^*) \right]
\nonumber \\
       &+& \frac{1}{2}\left\{ \Gamma_V \,(2\gamma\varrho_0 (k_F))^2 
       - \Gamma_S\, (2\varrho_{S}(M^*) )^2 \right\}
\label{pt}\\
p_\parallel &=& 2\left[ \frac{1}{4} E_F \varrho_0 (k_F) 
       - \frac{1}{4} M^* \,\varrho_{S}(M^*) 
       + E_F \varrho_0 (k_F)\, (\gamma^2 -1)  \right]
\nonumber \\
       &+& \frac{1}{2}\left\{ \Gamma_V \,(2\gamma\varrho_0 (k_F))^2 
       - \Gamma_S\, (2\varrho_{S}(M^*) )^2 \right\}
\label{pp}
\eeqa
To compare CNM with NM it is instructive to rewrite Eqs. (\ref{tm3}-\ref{pp}).  
The separation of projectile and target nucleons in momentum space 
increases the phase space volume in CNM. 
The boosted ellipsoids are elongated in longitudinal direction but in 
transverse direction the Fermi momenta $k_F$ 
of the subsystems (which are fixed by the current rest 
densities  $\varrho_0 (k_F)$) are not enhanced. 
This feature can be accounted for if we integrate over one Fermi 
sphere at rest $\Theta (k; k_F, u=0)$, however, with a doubled 
phase space occupancy factor $2\kappa$
\beq
<X >_{2\kappa}  = \frac{2\kappa}{(2\pi)^3} 
\int d^3 {\bf k} X(k) \Theta(k; k_F, u=0)~~.
\label{confint2}
\eeq
This description yields directly the CNM scalar density (\ref{rs1})
\beq 
\varrho_{S_{12}}  = < \frac{M^*}{E^*} >_{2\kappa} 
\label{scalar4}
\eeq
and $2\varrho_0 (k_F) = <1>_{2\kappa} $. One can now rewrite 
Eqs. (\ref{tm3}-\ref{pp}) separating thereby the static parts 
from those contributions which depend on the streaming velocity. 
Making use of (\ref{confint2}) the static parts of the energy and 
pressure densities are expressed as in ground state matter, however, 
with $\kappa =8$:  
\beqa
T^{00} &=& T^{00}|_{2\kappa} + 2 \gamma^2 {\bf u}^2 
\left[ E_F \varrho_0 (k_F)  + \Gamma_V \, \varrho_{0}^2 (k_F) \right]
\label{tm4}\\
p_\perp &=& p_{\perp}|_{2\kappa} 
+ 2 \gamma^2 {\bf u}^2 \,\Gamma_V \, \varrho_{0}^2 (k_F)
\label{pt2}\\
p_\parallel &=& p_{\parallel}|_{2\kappa}  + 2 \gamma^2 {\bf u}^2 
\left[ E_F \varrho_0 (k_F)  + \Gamma_V \, \varrho_{0}^2 (k_F)\right]
\label{pp2}
\eeqa
Eqs. (\ref{tm4}-\ref{pp2}) are exact as long as the ellipsoids do not 
overlap. It can be directly seen from there that CNM is characterized by two 
effects which act in opposite direction: The first one is the separation of 
the nucleons belonging to the different currents in phase space. This 
effect acts like an additional degree of freedom. Evidently it reduces the 
pressure of the Fermi motion. Since the scalar part of 
the potential energy is not affected by the spread in momentum space, 
besides the fact that the contributions of 
projectile and target are superposed, it deepens 
the scalar potential. However, the vector repulsion gives rise 
to an additional velocity dependent pressure in both, transverse 
and longitudinal direction (terms proportional to $\Gamma_V$). The 
elongation of the boosted Fermi ellipsoids 
in longitudinal direction enhances the baryon vector density which leads 
to an additional kinetic pressure in longitudinal direction 
($2 \gamma^2 {\bf u}^2 E_F \varrho_0 (k_F) )$. Both these contributions are 
related to the kinetic energy of the relative motion in the system 
and enhance the energy density. 

\begin{figure}[h]
\unitlength1cm
\begin{picture}(8.,6.0)
\put(-0.5,0.5){\makebox{\epsfig{file=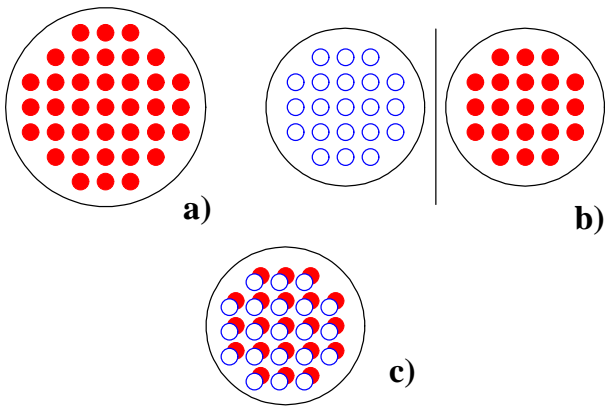,width=8.0cm}}}
\end{picture}
\caption{Schematic representation of the phase space in 
nuclear matter (a), colliding nuclear matter (b), and in 
colliding nuclear matter as experienced by transverse degrees of freedom, 
respectively by the scalar, density dependent part of the interaction (c).
}
\label{Fig2}
\end{figure}

The comparison to equilibrated NM should be performed at identical 
rest densities, i.e. the two currents with their rest 
densities $\varrho_{0} (k_F)$ should be 
compared to NM at the same total density
\beqa
\varrho_{NM}(k_{F_{tot}}) = 2\varrho_{0} (k_F)~~;~~
k_{F_{tot}} = 2^{\frac{1}{3}} k_F ~~.
\label{comp}
\eeqa
By this procedure one prevents a distortion of the comparison by the purely 
kinematical enhancement of the boosted subsystem densities 
in the c.m. frame which does not affect the transverse degrees of freedom. 
A comparison of CNM and NM at identical total baryon densities would mean to 
interpret the Lorentz contraction in the anisotropic system as a compression 
when related to the isotropic configuration. Condition (\ref{comp}) is 
also necessary for a meaningful comparison of CNM 
at different streaming velocities. Doing so, one can 
estimate the phase space effects comparing the contribution 
$T^{00}|_{2\kappa} (k_F)$ in (\ref{tm4}) with $T^{00}|_{NM} (k_{F_{tot}})$ in NM.  
Schematically, the separation of phase space in the colliding system 
is illustrated in Fig. \ref{Fig2}. 
%%%%%%%%%%%%%%%%%%%%%%%%%%%%%%%%%%%%%%%%%%%%%%%%%%%%%%%%%%%%%%%%%%%
%%%%%%%%%%%%%%%%%%%%%%%%%%%%%%%%%%%%%%%%%%%%%%%%%%%%%%%%%%%%%%%%%%%
It is clear that the introduction 
of a new degree of freedom, e.g. going from neutron matter to 
isospin saturated matter, strongly increases the binding energy. 
In contrast to the scalar field, where the colliding configuration can be 
exactly treated by the introduction of an additional degree of freedom 
(Fig.\ref{Fig2}a$\mapsto$c), the vector field 
is enhanced by the spread in momentum space 
(Fig.\ref{Fig2}b) and the elongation of the boosted ellipsoids 
(not shown in Fig.\ref{Fig2}b). However, the present picture is more 
general. Besides the fact that in CNM the internal 
Fermi pressure of the subsystems is reduced also the contributions of 
the {\it internal} potential energy  of the subsystems are modified due to 
the smaller Fermi momenta. The spread in momentum space, on the other 
hand, determines the interaction {\it between} the two currents and leads 
to additional repulsion in the combined system. These are the two 
essential features of colliding matter and - as will be seen later on - their 
implications on the EOS are relatively independent 
on the choice of the nuclear forces.

The above considerations (Eqs. (\ref{rs1}-\ref{pp2})) required 
that the ellipsoids were completely 
separated. The overlapping case which contains the limit 
$v_{\rm rel} \rightarrow 0$, i.e. equilibrated NM is more complicated. 
This involves the Pauli correction and the construction of the 
CNM configuration without double occupancy 
\cite{sehn96}. The principle of this procedure is to keep 
the streaming velocities $u_i$ fixed and to enlarge the corresponding 
Fermi momenta ($k_F \rightarrow {\tilde k}_F$) 
thus that the condition  
\beqa
j_{0\, 12} ({\tilde k}_F) &=& \frac{\kappa}{(2\pi)^3} \int d^3 k 
\left[ \Theta_1 (k;{\tilde k}_F, u) +\Theta_2 (k;{\tilde k}_F, -u) - 
\Theta_1 (k;{\tilde k}_F, u) \Theta_2 (k;{\tilde k}_F, -u) \right] 
 \nonumber \\
&=& 2\gamma(u) \varrho_0 (k_F)
\label{pauli1}
\eeqa
is fulfilled. Eq. (\ref{pauli1}) is valid for the symmetric case 
but can easily be extended to asymmetric configurations \cite{sehn96}. 
In the mean field approximation the integrals over the 
intersection ellipsoids can be worked out analytically which 
has been done by L. Sehn \cite{thesis}. For the energy 
density one ends up with 
\beqa
T^{00} &=& \frac{1}{4} M^*({\tilde k}_F) \,(2\varrho_{S}({\tilde k}_F) 
           -\delta\varrho_{S})  
\nonumber \\
       &+& \frac{1}{2}\left\{ \Gamma_V \,
       \gamma^2 (2\varrho_0({\tilde k}_F) -\delta\varrho_{0} )^2 
       + \Gamma_S\, (2\varrho_{S}({\tilde k}_F)-\delta\varrho_{S} )^2 \right\}
\nonumber \\
       &+& \frac{\kappa}{8\pi^2} \left\{ -\frac{1}{2}M^{*4} 
{\rm ln} \left[ \sqrt{\frac{1+u}{1-u}} \frac{{\tilde E}_F - {\tilde k}_F}{M^*}
         \right]\right.
\nonumber \\
       &-& \left. {\tilde k}_F {\tilde E}_F M^{*2} 
+ \frac{1}{3u}\gamma^2  {\tilde E}_F({\tilde E}_F-u{\tilde k}_F)^3 
-\frac{1}{3u}\left( \frac{{\tilde E}_F  }{\gamma}\right)^4\right.
 \nonumber \\
&-&  \left. \frac{1}{2}\gamma^2 M^{*2} (u{\tilde E}_F-{\tilde k}_F)
({\tilde E}_F+u{\tilde k}_F)\right\}~~.
\label{tm5}
\eeqa
The Pauli correction terms $\delta\varrho_{S}$ and $\delta\varrho_{0}$ 
(originating from the $\Theta_1 \Theta_2$ overlap integrals) 
can be found in \cite{sehn96}. 
In the limit $u\rightarrow 0$ the CNM configuration (\ref{conf}) approaches 
smoothly the corresponding NM configuration $\Theta (k_{F_{tot}})$ with 
${\tilde k}_F \rightarrow k_{F_{tot}}$, 
$\delta\varrho_{S}({\tilde k}_F)\rightarrow \varrho_{S}(k_{F_{tot}})$ and  
$\delta\varrho_{0}({\tilde k}_F)\rightarrow \varrho_{0}(k_{F_{tot}})$. The 
same holds for the energy momentum tensor, i.e. 
$T^{\mu\nu} ({\tilde k}_F, u\rightarrow 0)
\rightarrow T^{\mu\nu} (k_{F_{tot}},u=0)$. The enhancement factor 
which accounts in (\ref{rs1}-\ref{pp2}) effectively for the enlarged phase space 
volume approaches unity, i.e. $2\kappa \rightarrow\kappa$. 
%%%%%%%%%%%%%%%%%%%%%%%%%%%%%%%%%%%%%%%%%%%%%%%%%%%%%%%%%%%%%%%%%%%
\section{The effective EOS}
%%%%%%%%%%%%%%%%%%%%%%%%%%%%%%%%%%%%%%%%%%%%%%%%%%%%%%%%%%%%%%%%%%%
In colliding nuclear matter the energy per particle is defined in the usual way 
\beq
E_{12} (\varrho_{12},u) = {\rm T}^{00} / \varrho_{12} -M
\label{ekin}
\eeq
with $ \varrho_{12} = \sqrt{ j_{12}^2}$ the invariant 
baryon density in the c.m. frame of the two currents. 
%%%%%%%%%%
\footnote{According (\ref{pauli1}) the c.m. density is related to the 
subsystem rest densities by $\varrho_{12} =2\gamma(u) \varrho_0 (k_F)$. 
To avoid the stretching of the density scale by the Lorentz contraction 
we will in the following represent the energy  density as a function 
of the subsystem rest densities $2\varrho_0 (k_F)$. Physically this means 
to remove the Lorentz contraction in the representation of the free 
energy.}
%%%%%%%%%%
However, a meaningful discussion of non-equilibrium effects 
with respect to the ground state EOS should be based on 
the binding energy per particle. In the temperature zero case this 
corresponds to the free energy of the system where the 
contribution from the relative motion of the two currents has been 
subtracted. The corresponding effective EOS in colliding matter is 
then also directly linked to the hydrodynamical picture \cite{stoecker86}. 
The kinetic energy of a nucleon inside the medium is given by 
$E^{*}_{\rm kin} =  E^*(k) - M^*(k)$. Analogously, we identify the 
kinetic energy density of the relative motion in (\ref{tfull}) with 
\beqa
{\cal E}_{\rm kin} = < {\tilde k}^{\ast 0} (k^{\ast 0}-M^*) / {\tilde E}^{\ast} >_{12} 
=  < E^\ast (k) -M^* (k) >_{12}~~.
\label{ed1}
\eeqa
The effective mass $M^*$ accounts in (\ref{ed1}) for the 
fact that the two currents are interacting 
by momentum dependent forces. Such a momentum dependence occurs in the 
relativistic treatment already on the mean field level where the 
self-energies do not explicitely depend on momentum. The 
magnitude of $M^*$ characterizes thereby the strength of the momentum 
dependence of the optical potential.  
On the other hand side, different parameterizations of QHD including additional 
non-linear self-interaction terms for the scalar $\sigma$ field \cite{bog82} 
can yield identical EOSs but differ in their momentum dependence. This 
effect is taken into account in (\ref{ed1}). Moreover, 
the kinetic energy equals $E^\ast  -M^* \simeq \frac{{\bf k}^2}{2M^*}$ 
in the non-relativistic limit. In the non-relativistic 
limit $M^*$ plays the role of the Landau mass which is used to 
parameterize the optical potential. 

The energy of the relative motion of two interacting currents 
is more subtle to define. The present description is 
inspired by the fact that in the non-relativistic case this should be 
the energy which is necessary to shift the two 
Fermi spheres in their c.m. frame against each other. 
In the relativistic framework this corresponds to 
the difference of the kinetic energy in the combined system and 
the corresponding NM configuration at twice the 
subsystem rest density $2\varrho (k_F)$ 
\beq
{\cal E}_{\rm rel}(k_F,u) = < E^* - M^* >_{12} 
-  < E^* - M^* >_{u=0}~.
\label{ebind0}
\eeq
By definition ${\cal E}_{\rm rel}$
contains the kinetic energy which arises from the separation of the 
two distributions, respecting thereby the Pauli principle, and 
accounts for the interaction between the two currents by the 
presence of the effective mass. The vector density is not 
conserved when boosting the two distributions against each other. 
The Lorentz contraction contributes therefore 
to the energy of the relative motion. Similar as the 
c.m. energy density ($\epsilon = T^{00}|_{\rm c.m.} = u_\mu T^{\mu\nu}u_\nu$), 
Eq. (\ref{ebind0}) can be formulated in an invariant manner 
\beqa
 {\cal E}_{\rm rel}(k_F,u) &=&
  u_{12\,\mu} < {\tilde k}^{\ast \mu} (k^{\ast \nu}-M^* \delta^{\nu}_0 ) 
   / {\tilde E}^{\ast} >_{12}  u_{12\,\nu}
\nonumber\\
 && - u_{1\,\mu} < {\tilde k}^{\ast \mu} (k^{\ast \nu}-M^* \delta^{\nu}_0 ) 
   / {\tilde E}^{\ast} >_{1} u_{1\,\nu}~~.
\label{ebind1}
\eeqa
Now one obtains the binding energy per particle 
\beq
E_{12}^{\rm bind}(k_F,u) = 
\frac{ T^{00} -{\cal E}_{\rm rel}}{\varrho_{12}} -M
\label{ebind}
\eeq
as a function of the total c.m. density 
$\varrho_{12}$ and the c.m. streaming velocities. 
In the following we will only consider symmetric configurations 
($k_{F_{1,2}}=k_F,u_{1,2}=\pm u $). The basic features of the 
colliding configuration are easier discussed at the mean field 
level where formulas have been evaluated analytically and thus 
we consider this case first.
%%%%%%%%%%%%%%%%%%%%%%%%%%%%%%%%%%%%%%%%%%%%%%%%%%%%%%%%%%%%%%%%%%%%%%%%%%%%%
\subsection{Mean field approximation}
%%%%%%%%%%%%%%%%%%%%%%%%%%%%%%%%%%%%%%%%%%%%%%%%%%%%%%%%%%%%%%%%%%%%%%%%%%%%%
Here we mainly use the QHD-I version of the Walecka 
model with only two free parameters adjusted to the nuclear matter 
bulk properties. The model yields a binding energy of -15.75 MeV, a 
relatively large saturation density ($k_{F_{\rm sat}}=1.42~{\rm fm}^{-1}$), 
a small effective mass of $M^*/M$=0.556 and a large compression modulus of 
K=540 MeV. 
%%%%%%%%%%%%%%%%%%%%%%%%%%%%%%%%%%%%%%%%%%%%%%%%%%%%%%%%%%%%%%%%%%%%%%%%%%%%%% 
\begin{figure}[h]
\unitlength1cm
\begin{picture}(10.,9.0)
\put(-1.5,0.0){\makebox{\epsfig{file=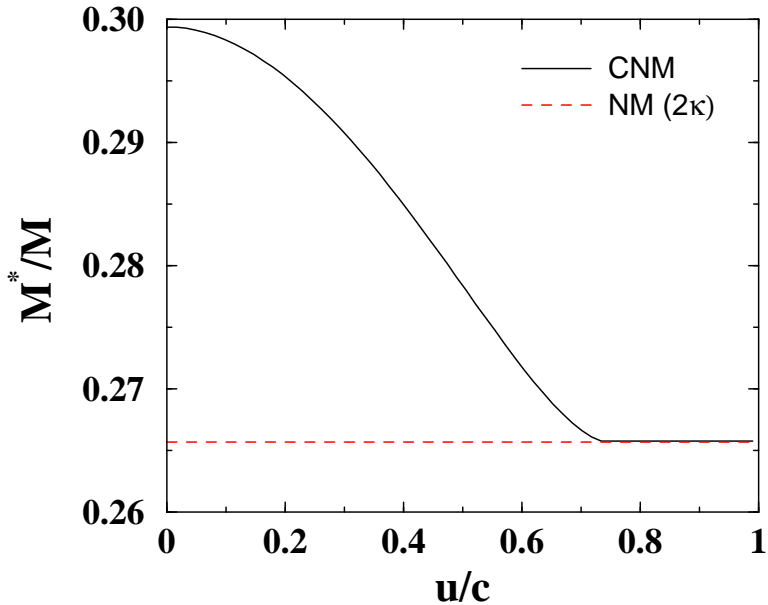,width=10.0cm}}}
\end{picture}
\caption{Effective mass in colliding nuclear matter (CNM) at 
subsystem densities $\varrho_0=\varrho_{\rm sat}$ as a function of the 
streaming velocity (solid) compared to the corresponding 
effective mass in nuclear matter obtained at $\varrho_0$ with 
a doubled phase space factor ($2\kappa$, dashed). 
Results are obtained with the Walecka model QHD-I.  
}
\label{Fig3}
\end{figure}
%%%%%%%%%%%%%%%%%%%%%%%%%%%%%%%%%%%%%%%%%%%%%%%%%%%%%%%%%%%%%%%%%%%%%%%%%%%%%
To illustrate the influence of the two-current geometry on the scalar 
part of the interaction in Fig. \ref{Fig3} the effective mass in 
colliding nuclear matter at 
subsystem densities $\varrho_{\rm sat}$  is shown as a function of the 
streaming velocity. At zero velocity the nuclear matter result 
at twice saturation density ($\varrho=2\varrho_{\rm sat})$ is obtained. 
With increasing streaming velocities the effective 
mass drops and reaches around $u/c=0.7$ (which corresponds to 
$E_{\rm lab}$=1.8 GeV) the asymptotic value of the non-overlapping configuration 
given by expression (\ref{rs1}). 
A further increase of $u$ has no influence on the effective mass since 
the scalar densities can be evaluated in the local rest frames of the 
currents. At the mean field level $M^*$ is exclusively 
determined by the shape of the momentum distribution, but once the ellipsoids 
are separated, the effective mass stays constant. The asymptotics are given  
by the corresponding effective mass in nuclear matter at $\varrho_0$ with 
the enlarged phase space factor ($2\kappa$). Hence the two-ellipsoid geometry 
enhances the attractive scalar field by about 30 MeV. 
However, in the total energy per particle (\ref{ekin}) this 
enhancement is completely compensated by the additional repulsive 
contributions in the counter-streaming system. As discussed above, Eqs. 
(\ref{tm4}-\ref{pp2}), an additional longitudinal Fermi pressure occurs 
due to the boosted vector densities. Secondly, the vector field gives 
rise to a strong repulsive force between the two currents.
%%%%%%%%%%%%%%%%%%%%%%%%%%%%%%%%%%%%%%%%%%%%%%%%%%%%%%%%%%%%%%%%%%%%%%%%%%%%%% 
\begin{figure}[h]
\unitlength1cm
\begin{picture}(14.,9.0)
\put(0.0,0.0){\makebox{\epsfig{file=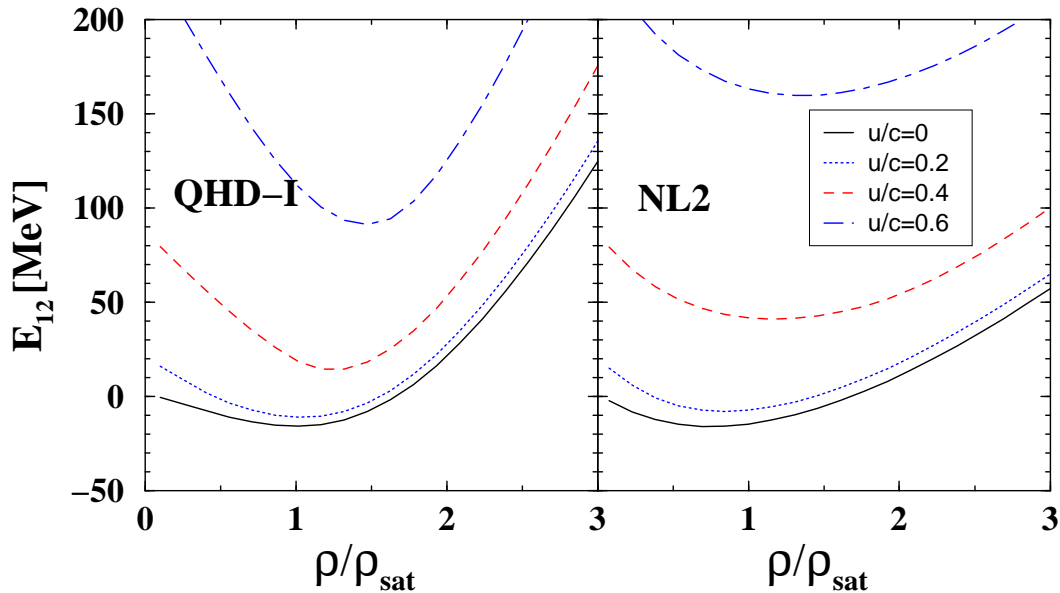,width=14.0cm}}}
\end{picture}
\caption{Energy per particle $E_{12}$ in colliding nuclear matter as 
a function of the subsystem density $2\rho_0/\varrho_{\rm sat}$ at 
different streaming velocities $u/c$=0/0.2/0.4/0.6. 
Results are obtained with the Walecka model QHD-I (left) and 
the non-linear Walecka model NL2 (right). In both cases densities are 
normalized to the saturation density $\varrho_{\rm sat}=0.193~{\rm fm}^{-3}$ 
of QHD-I. 
}
\label{Fig4}
\end{figure}
%%%%%%%%%%%%%%%%%%%%%%%%%%%%%%%%%%%%%%%%%%%%%%%%%%%%%%%%%%%%%%%%%%%%%%%%%%%%%
As can be seen from 
Fig. \ref{Fig4} the additional scalar attraction in the total energy per 
particle (\ref{ekin}) is then completely hidden by the much stronger 
repulsive contributions. Already at very low streaming velocities 
the system appears to be unbound. One should, 
however, be aware that QHD-I is a model with an unrealistically strong 
repulsive character at higher momenta, expressed by its small effective 
mass and the large vector field. As a result QHD-I strongly over predicts 
the empirical optical nucleon-nucleus potential \cite{opt} and the 
transverse nucleon flow in heavy ion collisions \cite{bkc88,fu96b}. 
To estimate the model dependence inherent in the present considerations 
in Fig. \ref{Fig4} we compare also to a softer version, the 
non-linear Walecka model NL2 \cite{wast88}, with bulk properties of 
$E= -15.75$ MeV, $k_{F_{\rm sat}}=1.29~{\rm fm}^{-1}$, K=237 MeV. 
A large effective mass $M^*/M$=0.80, respectively a weaker momentum 
dependence leads to a more realistic description of the dynamics 
in heavy ion reactions \cite{bkc88,fu96b}. Although less repulsive at 
large momenta the NL2 parameterization yields qualitatively the same 
velocity dependence of the total energy per particle as QHD-I. 
%%%%%%%%%%%%%%%%%%%%%%%%%%%%%%%%%%%%%%%%%%%%%%%%%%%%%%%%%%%%%%%%%%%%%%%%%%%%%% 
\begin{figure}[h]
\unitlength1cm
\begin{picture}(14.,9.0)
\put(0.0,0.0){\makebox{\epsfig{file=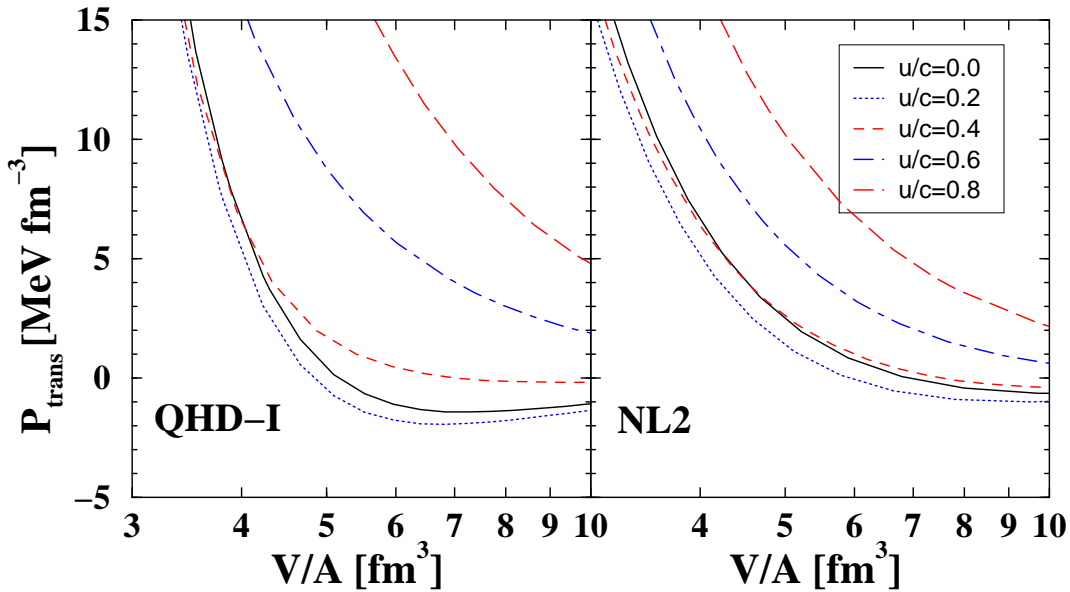,width=14.0cm}}}
\end{picture}
\caption{Transverse pressure $p_\perp$ in colliding nuclear matter as a 
function of the volume per particle at different streaming velocities. 
Results are obtained with the Walecka model QHD-I (left) and 
the non-linear Walecka model NL2 (right). 
}
\label{Fig4b}
\end{figure}
%%%%%%%%%%%%%%%%%%%%%%%%%%%%%%%%%%%%%%%%%%%%%%%%%%%%%%%%%%%%%%%%%%%%%%%%%%%%%
In Fig. \ref{Fig4b} the transverse 
pressure $p_\perp = T^{11}= T^{22}$ is shown as a 
function of the volume per particle $V/A = 1/\varrho_{12}$. Interestingly, 
the transverse pressure is reduced at low streaming velocities. 
Although the momentum distribution of the two currents 
is still quite compact at $u/c=0.2$ 
the reduction of the Fermi pressure 
and the reduction of the effective mass by the 
enlarged phase space volume are sufficient to produce this softening in 
transverse direction. However, with increasing velocity the vector repulsion 
starts to dominate. Qualitatively this behavior is model 
independent, i.e. it occurs in QHD-I and the softer NL2 model. 
Only at the highest velocity shown in Fig. \ref{Fig4b} ($u/c=0.8$) 
the ellipsoids are 
well separated, i.e. eq. (\ref{pt2}) is exactly valid. At smaller $u$ the 
Pauli corrections are taken into account. 

In order to investigate the subtle features of the colliding system 
we consider now the effective binding energy per particle (\ref{ebind}). 
With eq. (\ref{tm2}) the energy of the relative motion can easily be 
evaluated in mean field approximation (no overlap): 
\beqa
{\cal E}_{\rm rel} &=& 2\varrho_0 (k_F) \left[ E_{F} 
\left(\gamma^2 -\frac{1}{4}\right) 
 - \frac{3}{4}  E_{F_{\rm tot}} - \gamma M^* + M^* (k_{F_{\rm tot}}) \right]
\nonumber \\
 &+& \frac{1}{4} \left[ M^* \varrho_{S\,12} - M^* (k_{F_{\rm tot}}) 
\varrho_{S}(k_{F_{\rm tot}}) \right]~~.
\label{ek2}
\eeqa
The meaning of (\ref{ek2}) becomes more transparent in the low density 
limit. Using  
\beq
\varrho_S = \varrho_0 \left[ 1-\frac{3}{10}\left(\frac{k_F}{M^*}\right)^2 
+{\cal O} \left(\left(\frac{k_F}{M^*}\right)^4\right)\right] 
\eeq
one obtains after an expansion in $u/c$
\beqa
{\cal E}_{\rm rel} = 2\varrho_0 (k_F) \left[ 
\frac{ M^*}{2} u^2 + \frac{3}{5}\frac{k_{F}^2}{2  M^{*} } 
+ \frac{k_{F}^2}{2  M^*}u^2 
- \frac{3}{5} \frac{k_{F_{\rm tot}}^2}{2M^* (k_{F_{\rm tot}})} \right]
+{\cal O}(u^4)~~.
\label{ek3}
\eeqa
Now it becomes evident that eq. (\ref{ek3}) contains the kinetic energy 
of two freely streaming gases $T^{00}-\gamma M^* \varrho_0$. The first 
term is just the mean kinetic energy of a comoving nucleon, the second 
term is the kinetic energy of the internal Fermi motion in the local 
rest frame and the third contribution is due to the elongation of the 
Fermi ellipsoids in longitudinal direction by the boosts. Subtracted is 
the kinetic energy of the Fermi motion of the rest system at equal rest 
densities. Hence the definition (\ref{ek2}) respects the 
Pauli principle. An alternative choice would be to subtract simply the 
kinetic energy of the superimposed two spheres as illustrated by 
Fig. \ref{Fig2}c which would, however, violate the Pauli principle. 
To subtract the kinetic energy of the rest system at identical, Lorentz 
enhanced vector densities, on the other hand side, would introduce an 
additional unreasonable velocity dependence into the last term of eq. 
(\ref{ek3}). Thus eq. (\ref{ebind0}) is the natural definition of the 
relativistic kinetic energy of the relative motion and contains the correct 
non-relativistic limit.
%%%%%%%%%%%%%%%%%%%%%%%%%%%%%%%%%%%%%%%%%%%%%%%%%%%%%%%%%%%%%%%%%%%%%%%%%%%%%% 
\begin{figure}[h]
\unitlength1cm
\begin{picture}(14.,9.0)
\put(0.0,0.0){\makebox{\epsfig{file=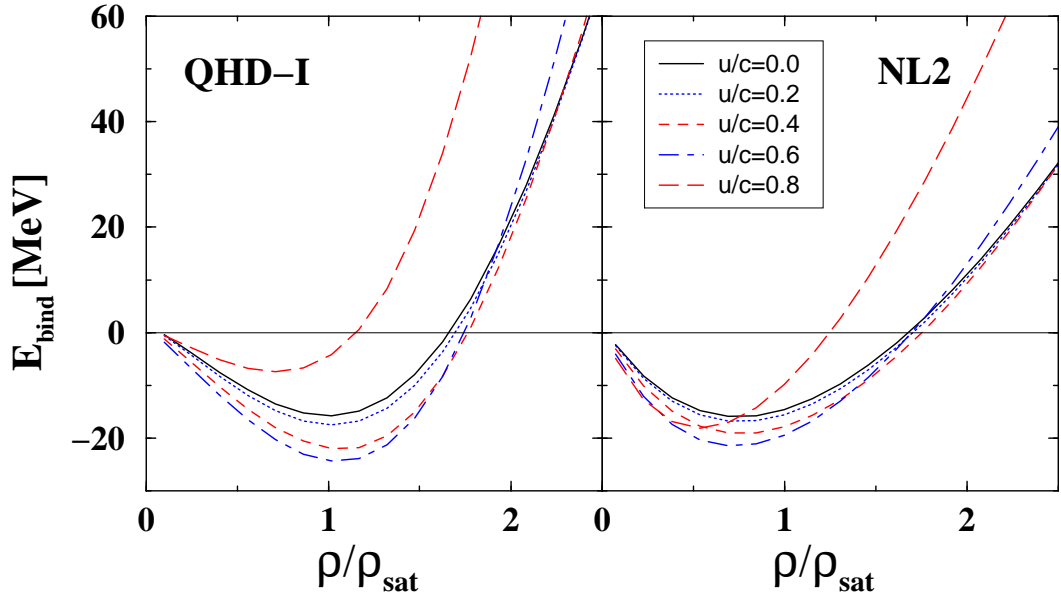,width=14.0cm}}}
\end{picture}
\caption{Effective binding energy per particle $E_{12}^{\rm bind}$ 
in colliding nuclear matter as a function of the subsystem 
density $2\rho_0/\varrho_{\rm sat}$ at different streaming velocities. 
Results are obtained with the Walecka model QHD-I (left) and 
the non-linear Walecka model NL2 (right). In both cases densities are 
normalized to the saturation density $\varrho_{\rm sat}$ of QHD-I. 
}
\label{Fig5}
\end{figure}
%%%%%%%%%%%%%%%%%%%%%%%%%%%%%%%%%%%%%%%%%%%%%%%%%%%%%%%%%%%%%%%%%%%%%%%%%%%%%
Now one is able to construct an effective EOS which relates colliding 
to ground state matter. In Fig. \ref{Fig5} 
the corresponding  effective equations of state are 
shown. First of all it is seen that the two-Fermi-ellipsoid 
geometry leads to a softening of the effective EOS at moderate streaming 
velocities. This can be understood by the reduced Fermi 
pressure in transverse direction 
and the enlarged scalar attraction as compared to NM. The vector 
potential is not affected by the subtraction scheme for ${\cal E}_{\rm rel}$ 
and contributes fully to $E_{12}^{\rm bind}$. 
Thus, the enhancement of the  vector repulsion 
acts in opposite direction and makes 
the EOS harder at large densities and/or high 
streaming velocities. Here the model dependences become more significant. 
At $u/c=0.8$ the effective QHD-I EOS lies well above the corresponding 
ground state result whereas for NL2 it is still softer at low densities. 
This reflects nicely the weaker momentum dependence and the weaker 
repulsion of NL2. In summary the softening of the ``'effective EOS''' by 
about 10 MeV due to the anisotropic phase space distributions in CNM is 
hardly seen on the scale of the total energy per particle 
but on the scale of an effective binding energy. 
Such small effects are, however, the features one is 
after in heavy ion experiments at intermediate energies.

%%%%%%%%%%%%%%%%%%%%%%%%%%%%%%%%%%%%%%%%%%%%%%%%%%%%%%%%%%%%%%%%%%%%%%%%%%%%%
\subsection{DB approximation}
\begin{figure}[h]
\unitlength1cm
\begin{picture}(14.,9.0)
\put(0.0,0.0){\makebox{\epsfig{file=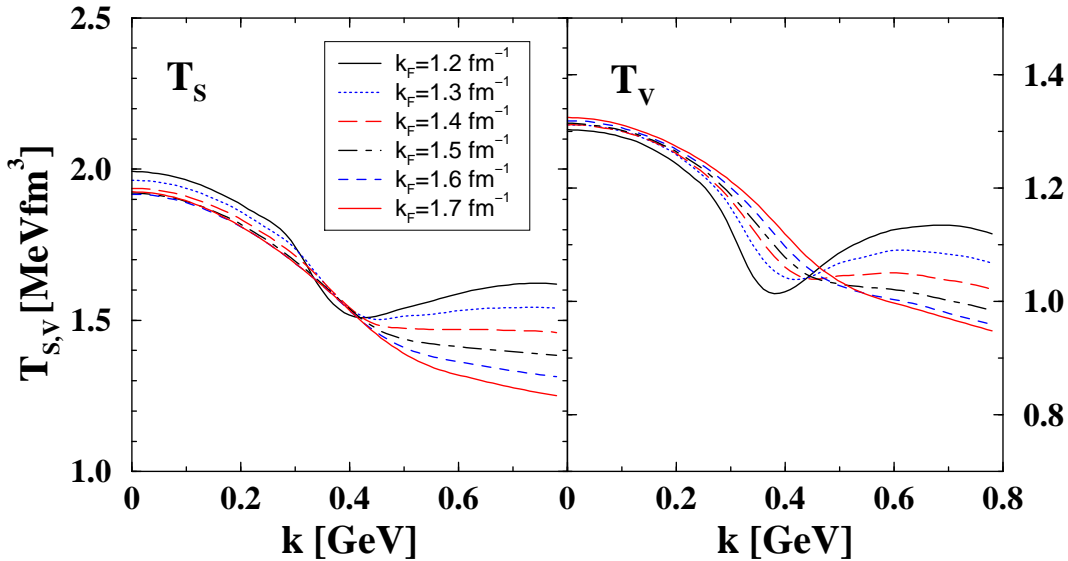,width=14.0cm}}}
\end{picture}
\caption{Averaged scalar and vector T-matrix amplitudes ${\bar T}_{\rm S,V}$ 
at various densities. The amplitudes are obtained from DB calculations 
for NM at rest. 
}
\label{Figtmat}
\end{figure}

%%%%%%%%%%%%%%%%%%%%%%%%%%%%%%%%%%%%%%%%%%%%%%%%%%%%%%%%%%%%%%%%%%%%%%%%%%%%%
The mean field used in the present section is  based on recent 
DB calculations \cite{boelting99} with nuclear matter 
saturation properties (Bonn A potential) 
of $\rho_{\rm sat}= 0.185~{\rm fm}^{-3},~E^{\rm bind} = -16.15$ MeV, 
a compression modulus of $K= 230$ MeV and an effective mass of 
${\tilde M}^*/M = 0.678$ which lies in between the phenomenological models  
QHD-I and NL2. In ground state matter the momentum 
dependence of the DB self-energies is 
moderate above the Fermi surface. 
The solution of the full DB problem for colliding nuclear 
matter, i.e. the self-consistent solution of the BS-equation for 
two-Fermi-ellipsoid configurations, is 
still an unresolved problem. Thus we determine the effective 
interaction in CNM as described in \cite{sehn96}. The 
ground state T-matrix amplitudes are parameterized in form of 
averaged momentum and density dependent scalar and vector coupling 
functions
\beqa
{\bar T}_S (k;k_F) &=& \int \frac{d^3{\bf q}}{\tilde{E}^*({\bf q})}
{\tilde M}^{*} T_S (k,q;k_F)~ \Theta (q;k_F)
~/~\int \frac{d^3{\bf q}}{\tilde{E}^*({\bf q})}{\tilde M}^{*}~\Theta (q;k_F)
\label{tt1}\\
{\bar T}_V (k;k_F) &=& \int \frac{d^3{\bf q}}{\tilde{E}^*({\bf q})}
T_V (k,q;k_F)~ \Theta (q;k_F)
~/~\int \frac{d^3{\bf q}}{\tilde{E}^*({\bf q})}~\Theta (q;k_F)~~.
\label{tt2}
\eeqa
The amplitudes $T_{\rm S,V}$ in eqs. (\ref{self1},\ref{et1}) should 
self-consistently be determined for colliding matter configurations. 
However, in the present work they are 
approximated by the corresponding NM amplitudes ${\bar T}_{\rm S,V}$, 
read in the subsystems at the respectively transformed momenta $k^\prime$. 
Details of this procedure can be found in  \cite{sehn96}. The 
treatment is not self-consistent 
on the level of the effective two-body interaction, e.g. the 
influence of the anisotropic momentum configuration on the 
Pauli operator for the intermediate states in the BS-equation 
is not taken into account, but it accounts for the correct 
geometry and the integration of the T-matrix amplitudes 
over this geometry. Fig.\ref{Figtmat} displays the momentum 
dependence of the ${\bar T}_{\rm S,V}$ amplitudes at various densities. 
At high momenta which exceed the range of the DB calculations we assume 
a $1/k$ asymptotics for the amplitudes.

%%%%%%%%%%%%%%%%%%%%%%%%%%%%%%%%%%%%%%%%%%%%%%%%%%%%%%%%%%%%%%%%%%%%%%%%%%%%%% 
%%%%%%%%%%%%%%%%%%%%%%%%%%%%%%%%%%%%%%%%%%%%%%%%%%%%%%%%%%%%%%%%%%%%%%%%%%%%%

\begin{figure}[h]
\unitlength1cm
\begin{picture}(10.,12.0)
\put(0.0,0.0){\makebox{\epsfig{file=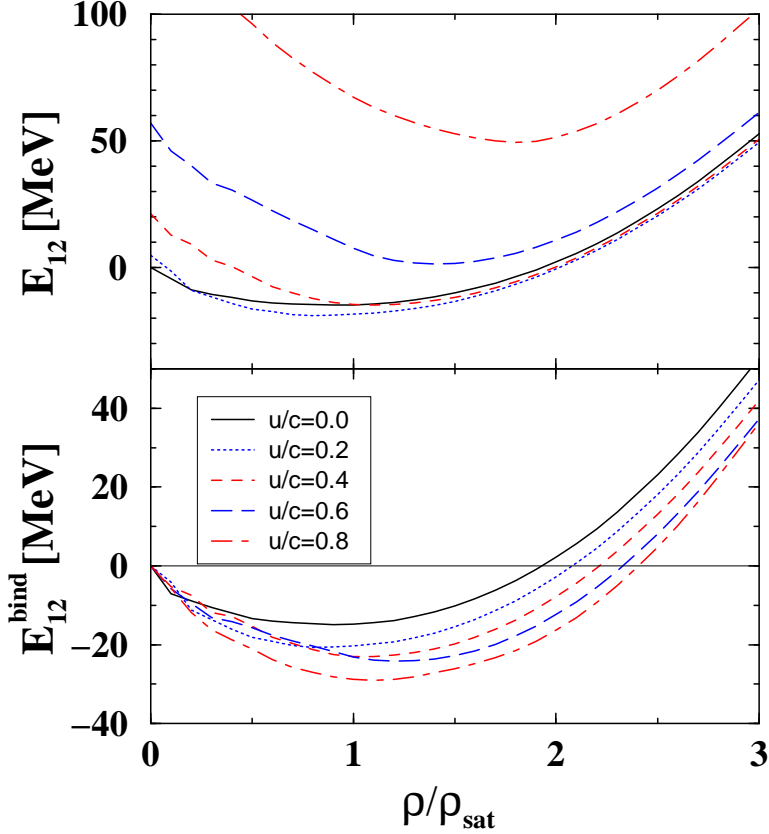,width=10.0cm}}}
\end{picture}
\caption{EOS in nuclear matter (solid) and colliding nuclear matter 
as a function of the subsystem 
density $2\rho_0/\varrho_{\rm sat}$ at different streaming velocities. 
The DB model is used. The upper and lower parts show the total energy per 
particle $E_{12}$ and the effective EOS, i.e. the 
binding energy per particle $E_{12}^{\rm bind}$ where the 
kinetic energy of the relative motion in CNM has been 
subtracted, respectively. The streaming velocities are 
$u$=0.2 (dotted), 0.4 (dashed), 0.6 (long-dashed), 
0.8 (dot-dashed). In both cases densities are 
normalized to the saturation density 
$\varrho_{\rm sat}$ the DB model. 
}
\label{Fig6}
\end{figure}

The corresponding equations-of-state are shown in Fig.\ref{Fig6}. The 
behavior of both, the energy per particle $E_{12}$ and the effective 
binding energy per particle $E_{12}^{\rm bind}$ are similar to the 
MF models discussed above. As already expressed by the saturation properties 
the DB model is significantly less repulsive than QHD-I. 
Moreover, the explicit density 
and momentum dependence of the amplitudes ${\bar T}_{\rm S,V}$ leads 
to some visible deviations from the MF picture. At low streaming velocities 
($u=0.2,~0.4$) even the total energy $E_{12}$ lies slightly below the 
ground state result. Consequently, the increase in binding energy seen 
in the effective EOS (lower part of  Fig.\ref{Fig6}) is more pronounced than 
in the MF approach, in particular at high densities. This behavior is due 
to the fact that the amplitudes  ${\bar T}_{\rm S,V}$ decrease with density 
and momentum. Compared to the MF case this momentum dependence 
leads to a reduction of both, the scalar and the vector field. 
In particular the reduction of the vector field, which increases in the 
MF models linearly 
with $\gamma$ and $\varrho$, results in considerably less repulsion for  
highly anisotropic CNM configurations. 

%%%%%%%%%%%%%%%%%%%%%%%%%%%%%%%%%%%%%%%%%%%%%%%%%%%%%%%%%%%%%%%%%%%%%%%%%%%%%% 
%%%%%%%%%%%%%%%%%%%%%%%%%%%%%%%%%%%%%%%%%%%%%%%%%%%%%%%%%%%%%%%%%%%%%%%%%%%%%

\begin{figure}[h]
\unitlength1cm
\begin{picture}(10.,9.0)
\put(-1.5,0.0){\makebox{\epsfig{file=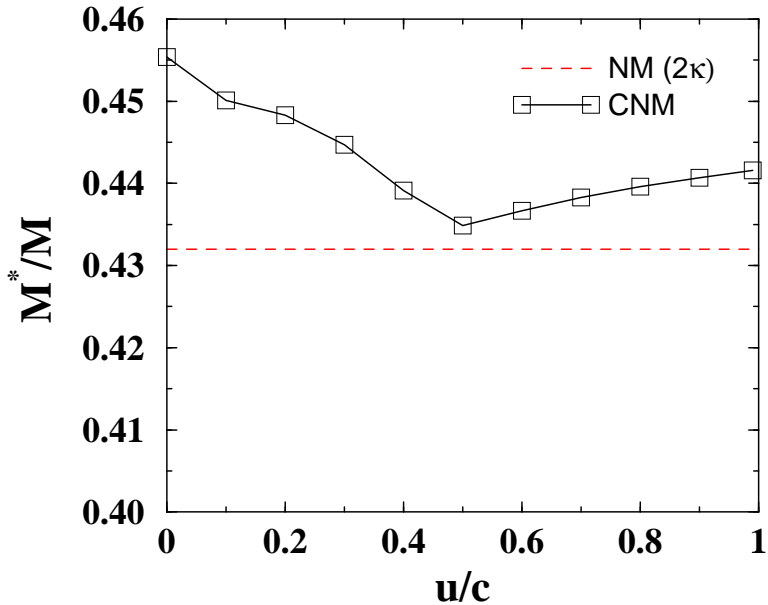,width=10.0cm}}}
\end{picture}
\caption{Effective mass in colliding nuclear matter (CNM) at 
subsystem densities $\varrho_0=\varrho_{\rm sat}$ as a function of the 
streaming velocity in the DB model. The result is compared to the corresponding 
effective mass in nuclear matter obtained at $\varrho_0$ with 
an enlarged phase space factor ($2\kappa$, dashed).
}
\label{Fig7}
\end{figure}

The momentum dependence of the T-matrix amplitudes 
is also reflected in the configuration dependence 
of the effective mass ${\tilde M}^*$ shown in Fig.\ref{Fig7} as a function  
of the streaming velocity at fixed subsystem densities 
$\varrho_0=\varrho_{\rm sat}$. As in the MF approximation, ${\tilde M}^*$ 
drops first with increasing anisotropy but does not fully reach the limit 
of a doubled phase space factor ($2\kappa$). Instead, ${\tilde M}^*$ 
starts to rise again at high relative velocities. This reflects the 
decrease of the scalar amplitude at large momenta, respectively the 
reduction of the scalar field when the momentum integration over the 
configuration is performed. In summary, the DB model follows 
to large extent the behavior of the 
MF approach. The leading kinematical effects such as the separation of 
phase space and a dropping effective mass as a function of the anisotropy 
are present. Superimposed are the density and 
momentum dependence of the T-matrix amplitudes. The explicit momentum 
dependence of the effective interaction leads to a reduced strength of 
both, the scalar and the vector field at large relative momenta which 
results in an even softer effective EOS in CNM compared to the MF picture. 

%%%%%%%%%%%%%%%%%%%%%%%%%%%%%%%%%%%%%%%%%%%%%%%%%%%%%%%%%%%%%%%%%%%%%%%%%%%%%% 
%%%%%%%%%%%%%%%%%%%%%%%%%%%%%%%%%%%%%%%%%%%%%%%%%%%%%%%%%%%%%%%%%%%%%%%%%%%%%

%%%%%%%%%%%%%%%%%%%%%%%%%%%%%%%%%%%%%%%%%%%%%%%%%%%%%%%%%%%%%%%%%%%%%%%%%%%%%
\section{Heavy ion reactions}
%%%%%%%%%%%%%%%%%%%%%%%%%%%%%%%%%%%%%%%%%%%%%%%%%%%%%%%%%%%%%%%%%%%%%%%%%%%%%
In this section the connection of the 
previous considerations with transport calculations is discussed. 
The early and high density phase of relativistic heavy ion reactions 
in the SIS ($E_{\rm lab}<$ 2 AGeV) and AGS  ($E_{\rm lab}<$ 10 AGeV) 
energy range are characterized by a high anisotropy of the local 
momentum space in beam direction. This has also practical implications for 
heavy ion reactions and the interpretation of experimental observables. 
In the overlapping zone of two interpenetrating 
nuclei the initial configuration is that of two sharp momentum 
ellipsoids. In the course of the reaction the mid-rapidity region 
is then more and more populated due to binary collisions and the system 
is heated up, i.e. the originally cold and sharp momentum 
ellipsoids become diffuse and merge together. At least in the central 
cell finally a fully equilibrated spherical configuration can be reached. This 
has e.g. been demonstrated in \cite{essler97,lang,puri}. 
%%%%%%%%%%%%%%%%%%%%%%%%%%%%%%%%%%%%%%%%%%%%%%%%%%%%%%%%%%%%%%%%%%%%%%%%%%%%%% 
\begin{figure}[h]
\unitlength 1cm
\begin{picture}(14.,9.0)
\put(0.0,0.0){\makebox{\epsfig{file=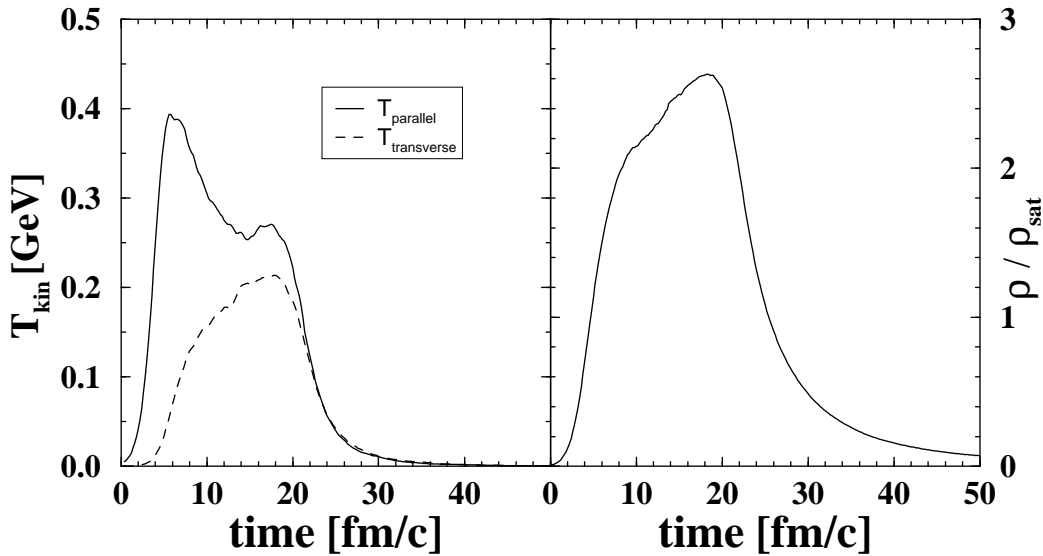,width=14.0cm}}}
\end{picture}
\caption{Time evolution of the kinetic part of the 
energy-momentum-tensor $T_{\rm kin}$ (left) and 
the baryon density (right) at the collision center in a 
central $Au+Au$ reaction at 1 A.GeV. The tensor components 
parallel $T_{\rm kin}^{\protect\parallel}$ and transverse 
$T_{\rm kin}^{\protect\perp}$ to the beam direction are 
shown separately.}
\label{Fig_tkin}
\end{figure}
%%%%%%%%%%%%%%%%%%%%%%%%%%%%%%%%%%%%%%%%%%%%%%%%%%%%%%%%%%%%%%%%%%%%%%%%%%%%%%
To obtain a more quantitative 
measure for the size and relevant time scales 
for phase space anisotropies in Fig.\ref{Fig_tkin} the time 
evolution of the kinetic part of the energy-momentum-tensor   
at the collision center of a central (b=0 fm) $Au+Au$ 
reaction at 1 A.GeV is compared to the 
time evolution of the corresponding baryon density. The 
difference between the parallel $T_{\rm kin}^{\parallel}= T_{\rm kin}^{33}$ 
and the transverse  
$T_{\rm kin}^{\perp}= (T_{\rm kin}^{11}+T_{\rm kin}^{22})/2$ 
components is thereby a measure for the anisotropy of the local momentum 
space in beam ($z$) direction. The quantities 
are obtained from relativistic BUU calculations where 
$T_{\rm kin}^{\mu\nu}$ is determined from the phase space distribution 
$f ({\bf x},{\bf k},t)$ represented by testparticles \cite{lca}
\beqa
T_{\rm kin}^{\mu\nu} ({\bf x},t)= 
< k^{\ast \mu} k^{\ast\nu} / E^{\ast} >_{f}  
= \frac{1}{N} \sum_{i}^{A\cdot N} u^{\mu}_i u^{\nu}_i m^{*}_i 
g({\bf x},{\bf x}_i,t) ~.
\label{tkinrlv}
\eeqa
The index $i$ in Eq.(\ref{tkinrlv}) refers to the testparticle and 
$g$ is a covariant Gaussian \cite{lca}. 
In the initial phase up to about 10 fm/c the local phase 
space is highly anisotropic, i.e. $T_{\rm kin}^{\parallel} >>T_{\rm kin}^{\perp}$. 
During the high density phase from $\sim~5\div 25$ fm/c the 
system starts to equilibrate. At the 
end of the high density phase and during the expansion 
phase the two components of the energy-momentum tensor are 
almost equal which indicates that a large amount of equilibration 
has been reached. 
Fig.\ref{Fig_tkin} demonstrates also that the relaxation time to 
reach equilibrium configurations 
($T_{\rm kin}^{\parallel} = T_{\rm kin}^{\perp}$) coincides 
more or less with the high density phase of the reaction.

\begin{figure}[h]
\unitlength1cm
\begin{picture}(10.,9.0)
\put(-1.5,0.0){\makebox{\epsfig{file=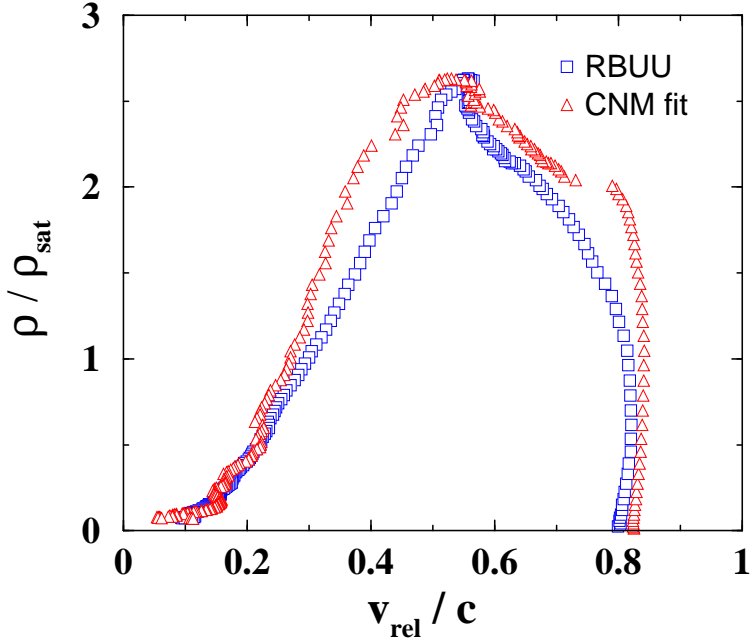,width=10.0cm}}}
\end{picture}
\caption{Time evolution of the two parameters, total density
 $\varrho_{0}$ and relative streaming velocity  $v_{\rm rel}$, which 
characterize anisotropic momentum configurations. The squares denote 
the values obtained in the central cell from a RBUU calculation for 
a central $Au+Au$ reaction at 1 AGeV. The triangles are 
the values determined by adjusting colliding nuclear matter configurations 
to the kinetic energy distributions of the RBUU calculation. Each symbol 
corresponds to a time step of 0.2 fm/c.
}
\label{Figvr}
\end{figure}

Now the question arises how accurately the BUU phase space can 
be reproduced by a sequence of colliding nuclear matter configurations. 
It is clear that the considered CNM configurations are an idealization 
of the real phase space. Firstly, one assumes symmetric 
configurations in projectile and target currents whereas the transport 
simulations contain density fluctuations. Secondly, the system is 
heated up whereas temperature effects are neglected in the cold 
configurations. To estimate the validity of this approximation 
the following procedure is applied: 

The temporal evolution of the two parameters 
$\varrho_0 = \varrho_{0}^T + \varrho_{0}^P$ 
and $v_{\rm rel}$ which characterize the anisotropic momentum space 
in terms of CNM are directly 
determined from the RBUU simulation, again in the central cell for 
$Au+Au$ at 1 AGeV. The effective compression density $\varrho_{0}$ 
is given as the sum 
of the projectile and target invariant rest densities 
$\varrho_{0}^{T/P}=\sqrt{j^{T/P}_{\mu}j^{\mu T/P}}$ and the 
velocity  $v_{\rm rel}= |{\bf v}_{\rm rel}|$ is defined via the 
projectile and target streaming velocities 
\beq
 {\bf v}_{\rm rel} = 
\frac{ {\bf u}^P - {\bf u}^T}{1- {\bf u}^P \cdot {\bf u}^T}~~.
\label{vrel}
\eeq
These values are compared to the corresponding $\varrho_{0}, v_{\rm rel}$ 
parameters for idealized symmetric ($\varrho_{0}^T = \varrho_{0}^P$) 
CNM configurations. Here $v_{\rm rel}$ is 
chosen in such a way that at given $\varrho_{0}$ the kinetic 
energy-momentum tensor components $T_{\rm kin}^{\parallel},T_{\rm kin}^{\perp}$ 
from the transport simulations are reproduced. The comparison is 
performed in Fig. \ref{Figvr} where 
each symbol corresponds to a time step of 
$\Delta t=0.2$ fm/c. Although not perfect, the CNM approximation 
is able to describe the time evolution in the $\varrho_{0}-v_{\rm rel}$ 
parameter space quite well. Since the CNM parameters are adjusted to 
reproduce the kinetic energy, deviations in the $v_{\rm rel}$ parameter 
can be regarded a measure for the difference between the real and the idealized 
configurations. These are due to temperature and finite size 
effects as well as to asymmetries arising from density fluctuations. One 
has thereby, however, to keep in mind that a description 
in terms of a naive local density approximation would completely neglect 
the dependence on the $v_{\rm rel}$ parameter.

From the time evolution of the system in the $\varrho_{0}-v_{\rm rel}$ 
parameter space  the following can be seen: Asymptotically 
the two nuclei start to touch at low density and high $v_{\rm rel}$. 
Then the density increases rapidly and reaches a maximum value of 
about 2.5 $\varrho_{\rm sat}$. In the beginning $v_{\rm rel}$ is large 
but then equilibration sets in. When maximal compression is 
reached the relative velocity is still relative high 
($v_{\rm rel}/c\simeq 0.6$), i.e. the local momentum space 
is still highly anisotropic. In the expansion phase 
the density drops rapidly and the system becomes more and more isotropic. 
Interestingly, the system stays for some time in the region of highest 
compression before it expands, i.e. it 
takes some time before an isotropic pressure builds up. Also in \cite{lang}
it was found that only in the final expansion phase the pressure is 
isotropic. Thus the momentum space anisotropy is most distinct before 
maximal compression is reached, consistent with Fig. \ref{Fig_tkin}. 
One effect which should also be noted is the slight increase of 
$v_{\rm rel}$ in the first two fm/c. This acceleration is a typical 
effect due to the interaction between the two currents. In the interacting 
system the effective mass drops which leads to the slight acceleration 
before the stopping due to binary collisions sets in. 
%%%%%%%%%%%%%%%%%%%%%%%%%%%%%%%%%%%%%%%%%%%%%%%%%%%%%%%%%%%%%%%%%%%%%%%%%%%%%% 
%%%%%%%%%%%%%%%%%%%%%%%%%%%%%%%%%%%%%%%%%%%%%%%%%%%%%%%%%%%%%%%%%%%%%%%%%%%%%
Thus one sees that the picture of two counter-streaming fluids 
is appropriate for a major part of the reaction, in particular when 
supra-normal densities are present. Certainly the assumption of 
cold fluids has to be improved since the system is heated up. However, 
with respect to the local density approximation where the $v_{\rm rel}$ 
dependence is to large extent neglected in the determination of the 
mean field the two-fluid picture provides an important step forward 
towards a more precise dependence of the mean field on the phase-space 
configuration. In transport calculations like BUU \cite{dani00} 
or QMD \cite{Ai91} or their relativistic counterparts RBUU \cite{bkc88,lca} 
and RQMD \cite{fu96b} the mean field is usually determined in 
a sort of a local density approximation. In the simplest approximation 
where no explicit momentum dependent forces are considered 
the fields depend only on the total density. In a non-relativistic approach 
this yields a local potential of the Skyrme type  
$U_{\rm s.p.} = \alpha \rho_B + \beta (\rho_{B})^\gamma$. In 
the relativistic case the single particle potential has the form  
\beqa
U_{\rm s.p.} (\rho_B, k) = k^0 -\sqrt{M^2 +{\bf k}^2} 
 = \sqrt{{\bf k}^{*2} + M^{*2}} - \Sigma_0 -\sqrt{M^2 +{\bf k}^2} ~~.
\label{upot1}
\eeqa
In mean field approximation, i.e. for Walecka type models, the 
scalar and vector fields depend 
on the local density, i.e. $M^* = M+\Sigma_S (\rho_B),~
\Sigma_0 = \Sigma_0 (\rho_B)$ and the lowest order momentum dependence of $U_{\rm s.p.}$, 
which is absent in non-relativistic treatments, originates from the difference 
between $M$ and $M^*$. 
It is worthwhile to extract the error which is made when the local density 
approximation (LDA) is applied to anisotropic momentum space configurations. 
Naturally, the LDA accounts only insufficiently for such a scenario. 
In transport simulation the particles are usually 
propagated in the center-of-mass frame of the colliding nuclei which 
corresponds to the c.m. frame of the two counter-streaming currents. 
There the baryon density is given by 
$\rho_B = \varrho_{12} = 2\gamma \varrho_0$ which means that the LDA 
treats the Lorentz contraction of the nuclei like a 
compression. 

To estimate the resulting error in the dynamics we compare 
the correct single particle potential for CNM 
to that obtained by making use of the LDA. Therefore a typical 
nucleon comoving with one of the currents with its average 
velocity $u$ is chosen. In the c.m. frame 
the  spatial components of the total current $j_{12}$ vanish and 
the vector field contains only the time like 
component. Hence the momentum of such a ``test'' nucleon is given by 
${\bf k}={\bf k}^* = M^* \gamma u {\hat e}_z$. Since the vector 
fields is on the mean field level identical in both, the CNM 
calculation and using the LDA, i.e. 
$\Sigma_0  =-\Gamma_V \varrho_{12} = -\Gamma_V \rho_{B}$, differences 
in the single particle potential 
arise only due the different values of the effective mass.

%%%%%%%%%%%%%%%%%%%%%%%%%%%%%%%%%%%%%%%%%%%%%%%%%%%%%%%%%%%%%%%%%%%%%%%%%%%%%% 
\begin{figure}[h]
\unitlength1cm
\begin{picture}(14.,9.0)
\put(0.0,0.0){\makebox{\epsfig{file=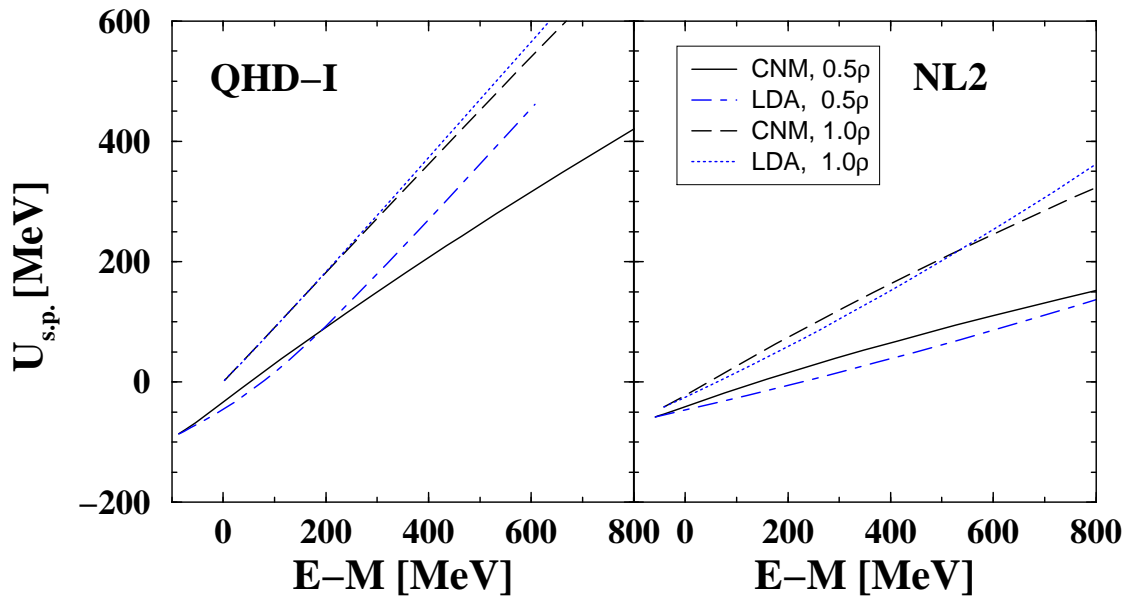,width=14.0cm}}}
\end{picture}
\caption{Single particle potential in colliding nuclear matter 
for a test nucleon which is comoving with one of the currents at 
subsystem rest densities of 0.5 and 1.0 $\varrho_{\rm sat}$. The 
CNM calculation is compared to the local density approximation. 
Results are obtained with the Walecka model QHD-I (left) and 
the non-linear Walecka model NL2 (right). 
}
\label{FigU1}
\end{figure}
%%%%%%%%%%%%%%%%%%%%%%%%%%%%%%%%%%%%%%%%%%%%%%%%%%%%%%%%%%%%%%%%%%%%%%%%%%%%%
In Fig. \ref{FigU1} the single particle potential (\ref{upot1}) for such 
a test nucleon moving with streaming velocity $u$ is shown as a function of the single 
particle energy $E_{\rm s.p.} = k^0 -M$,  for the two subsystem densities 
$\varrho_{0} = 0.5 \varrho_{\rm sat}~ {\rm and }~1.0 \varrho_{\rm sat} $. 
In this representation both, $E_{\rm s.p.}$ and the baryon density 
$\rho_B = 2\gamma \varrho_0$ depend on $u$. The effective mass and the 
resulting nucleon momentum are consistently determined for the colliding 
configuration (including Pauli effects). 
We compare this to the LDA where the fields are determined 
as for ground state nuclear matter at $ \rho_B$. Due to the 
different value for $M^*$ one obtains a different 
test nucleon momentum at fixed $u$. As seen from Fig.\ref{FigU1} 
the LDA leads to significant deviations in the single particle potential. 
That the deviations are even more pronounced at lower densities ($0.5\varrho_{\rm sat}$) 
comes from the fact that at high densities the potential is dominated by the 
vector field which coincides in the two approaches. Comparing QHD-I and NL2 
illustrates the model dependence of the error made by the LDA. At higher 
energies the LDA predicts generally a too repulsive potential, at low energies 
on the other hand too much attraction. The reason lies in the interplay 
between $M^*$ and $k$ which both enter into $U_{\rm s.p.}$ and $E_{\rm s.p.}$. 
For increasing streaming velocities at fixed subsystem rest 
densities the effective mass cannot drop below the 
asymptotic value reached for the completely  separated 
ellipsoids. However, the effective mass for ground state matter, 
determined as a function of the increasing baryon vector density, 
decreases below this limit. This leads to an 
overprediction of the scalar attraction in CNM and a correspondingly smaller 
value of the attributed momentum $k$. The interplay between these two effects 
creates the differences in the mean field potential.

A correct determination of the effective mass, respectively the scalar density 
is an important issue, not only with respect to the nucleon potential. 
Also mesons like pions, kaons, $\rho$-mesons ect. are assumed to change 
their properties in a dense nuclear environment \cite{brown96}. 
For the pseudo-scalar octet the Gell-Mann-Oakes-Renner relation \cite{gor}
connects the vacuum mass with the expectation value of the scalar quark 
condensate. Within the framework of QCD inspired effective models, like 
NJL \cite{njl}, effective chiral models \cite{kaplan86} etc. the change 
of the scalar quark condensate in the medium is usually expressed in terms of 
the scalar baryon density. E.g. the resulting in-medium dispersion relation for 
kaons \cite{li95,schaffner97,fuchs98}
\begin{equation}
\omega_{\rm{K^\pm}} ({\bf k}) = \sqrt{ {\bf k}^2 
+ m_{\rm K}^2 - \frac{\Sigma_{KN}}{ f_{\pi}^2} \varrho_S + V_{0}^2 } 
\pm V_0 
\label{disp2}
\end{equation}
contains an attractive scalar part $\Sigma_{KN} / (f_{\pi}^2) \varrho_S$ 
and the vector potential $V_\mu = 3/(8  f_{\pi}^2) j_\mu$ which is of 
different sign for kaons and antikaons. In particular kaons are primordially 
produced in the early and high density phase of the collisions \cite{li95} 
which means that they are born into an environment which 
resembles to large extent colliding matter. Already at SIS energies but particularly 
at AGS energies \cite{ko2000} the scalar density can be given by the asymptotic value 
(\ref{rs1}) for the corresponding CNM configuration, characterized by high 
streaming velocities, large $\gamma$-factors and high $\rho_B$. Hence the LDA 
will lead to an overestimation of the scalar density and the corresponding 
mass reduction of these particles. 
%%%%%%%%%%%%%%%%%%%%%%%%%%%%%%%%%%%%%%%%%%%%%%%%%%%%%%%%%%%%%%%%%%%%%%%%%%%%%% 
\begin{figure}[h]
\unitlength1cm
\begin{picture}(10.,9.0)
\put(-1.5,0.0){\makebox{\epsfig{file=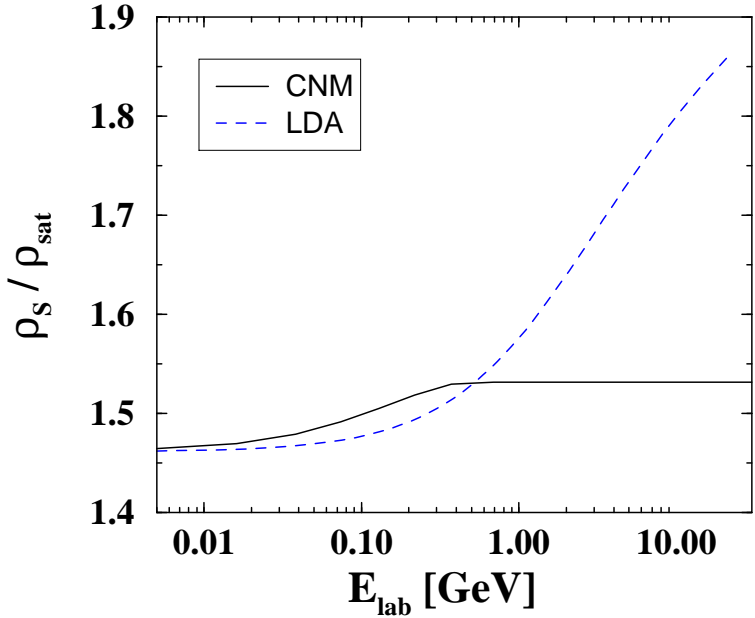,width=10.0cm}}}
\end{picture}
\caption{Scalar density in colliding nuclear matter (CNM) at 
subsystem densities $\varrho_0=\varrho_{\rm sat}$ as a function of the 
laboratory energy determined in the Walecka Model QHD-I. 
The result is compared to the corresponding 
effective mass in nuclear matter obtained in a local density approximation 
to the colliding system. 
}
\label{Figrs}
\end{figure}
%%%%%%%%%%%%%%%%%%%%%%%%%%%%%%%%%%%%%%%%%%%%%%%%%%%%%%%%%%%%%%%%%%%%%%%%%%%%%
To estimate this effect in Fig.\ref{Figrs} we show the scalar density 
in CNM as a function of the laboratory energy and compare to the LDA result.
The laboratory energy of the asymptotically non-interacting system is given by 
$E_{\rm lab} = M(\gamma(v)-1) = 2M(\gamma^2 (u)-1)$ where $v$ is the 
laboratory velocity and $u$ is again the c.m. streaming velocity. 
The subsystem rest densities are again chosen 
as $\varrho_0 = \varrho_{\rm sat}$ and QHD-I is used. 
The system is not compressed above $ 2\varrho_0$, i.e. the 
increase of $\rho_B =2\gamma (u)\varrho_0$ is only due to 
the relative motion of the two currents. In the LDA the 
Lorentz contraction is interpreted as compression 
and the scalar density is calculated 
at the correspondingly enlarged Fermi momentum. 
Evidently this standard procedure 
yields completely different results.

Consequences for observables, in particular collective 
flow observables, have been discussed in \cite{lca}. The considered 
non-equilibrium effects reduce the transverse flow which is consistent 
with the softening of the effective EOS discussed in Sec. III. The 
magnitude of this effect was found to be of the order as the usage 
of different (soft/hard) mean field parameterizations. 
%%%%%%%%%%%%%%%%%%%%%%%%%%%%%%%%%%%%%%%%%%%%%%%%%%%%%%%%%%%%%%%%%%%%%%%%%%%%%% 
\begin{figure}[h]
\unitlength1cm
\begin{picture}(10.,9.0)
\put(-1.5,0.0){\makebox{\epsfig{file=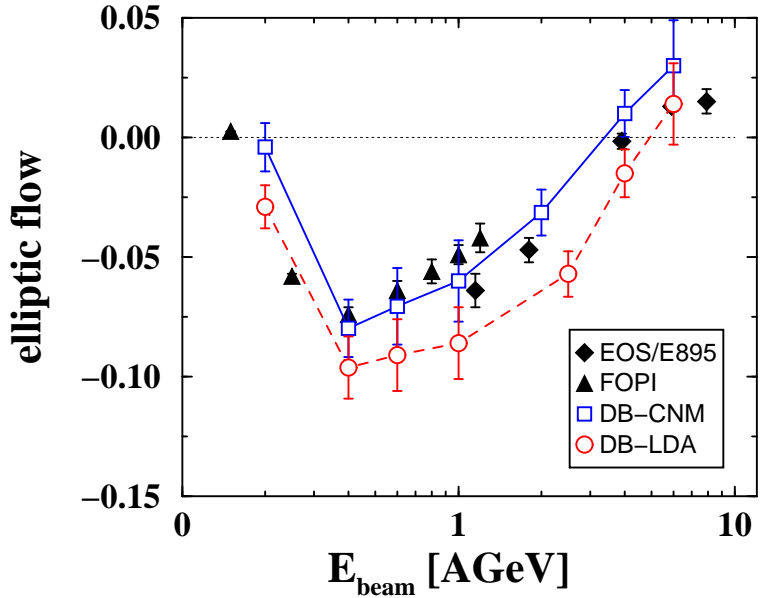,width=10.0cm}}}
\end{picture}
\caption{Excitation function of the elliptic flow $v_{2}$ in 
$Au+Au$ reactions at mid-rapidity. The data are 
taken from the FOPI- (triangles) \protect\cite{andronic99}, 
and EOS/E895 (diamonds) \protect\cite{dani00} collaborations. 
RBUU calculations are 
performed with DB forces applied in the local density approximation 
(LDA) and determing the mean field locally in the colliding nuclear 
matter approximation (CNM).
}
\label{Figflow}
\end{figure}
%%%%%%%%%%%%%%%%%%%%%%%%%%%%%%%%%%%%%%%%%%%%%%%%%%%%%%%%%%%%%%%%%%%%%%%%%%%%%
As an example we show in Fig. \ref{Figflow} the excitation function 
of the elliptic flow $v_2$ for $Au+Au$ reactions at mid-rapidity. 
The mean field in the corresponding RBUU calculations \cite{gaitanos01} 
is based on the DB model discussed in Sec. III. In the one case the 
mean field is simply treated in the local density approximation (LDA), 
in the second case the potentials for CNM were used. Here the local 
parameters $\varrho_{0}^{T/P} ({\bf x},t)$ and $v_{\rm rel} ({\bf x},t)$ 
are determined from the actual phase space distribution 
$f({\bf x},{\bf k},t)$. Details of the calculations can be found in 
\cite{lca} and \cite{gaitanos01}. The two different treatments which 
are based on identical nuclear forces yield significantly 
different results for the $v_2$. Since $v_2$ at mid-rapidity 
is in particular sensitive to the EOS at 
supra-normal densities \cite{eflow,dani00,gaitanos01} these effects 
are here most pronounced. 
Using the local density approximation one would exclude the underlying 
EOS from the comparison to data as too stiff. The more 
consistent treatment of momentum space anisotropies 
leads to the discussed net softening of the 
effective EOS in the course of the reaction and restores the agreement 
with experimental data. Therefore non-equilibrium effects should be 
taken into account on the level of the effective in-medium 
interaction which means to determine the mean field used in 
transport calculations consistently for colliding nuclear matter. 

%%%%%%%%%%%%%%%%%%%%%%%%%%%%%%%%%%%%%%%%%%%%%%%%%%%%%%%%%%%%%%%%%%%%%%%%%%%%%
\section{Non-relativistic approaches}
%%%%%%%%%%%%%%%%%%%%%%%%%%%%%%%%%%%%%%%%%%%%%%%%%%%%%%%%%%%%%%%%%%%%%%%%%%%%%
Since non-relativistic mean fields are also widely used in transport 
models it is worthwhile to shortly consider this case, in particular 
with respect to the structure of the single particle potential. 
For this purpose we choose a simple interaction which can serve as 
an illustrative example. The non-relativistic problem of two 
Galilei transformed 
Fermi spheres has e.g. been 
treated in \cite{neise90}. Using a spin-isospin averaged Skyrme force 
of the type 
\beqa
V({\bf k},{\bf q};{\bf k}^\prime,{\bf q}^\prime) 
= t_0 + \frac{t_1}{2}\left[({\bf k}- {\bf q})^2 + 
({\bf k}^\prime - {\bf q}^\prime)^2 \right]
+ t_2 ({\bf k}- {\bf q})\cdot ({\bf k}^\prime - {\bf q}^\prime)
\label{sk1}
\eeqa
the energy density in nuclear matter reads
\beqa
\epsilon = \frac{\tau}{2M} + \pi =  \frac{\tau}{2M} + 
\frac{1}{2}t_0 \varrho^2  + (t_1 + t_2) \varrho \tau ~~~.
\label{sk2}
\eeqa
Three-body forces which give rise to a quadratic density 
dependence of the mean field have been neglected in the 
interaction (\ref{sk1}). Thus we restrict the present discussion to 
the two-body level. The kinetic part of the energy density is given by 
\beqa
\tau = \frac{\kappa }{(2\pi)^3}\int d^3 k~{\bf k}^2 \Theta (k,k_F)
= \frac{2}{5\pi^2} k_{F}^5
\label{sk3}
\eeqa
with $\kappa =4 $ in spin-isospin saturated matter. The local mean field 
and the effective Landau mass are given by 
\beqa 
U_{\rm loc}(k_F) = \frac{\partial \epsilon}{\partial\varrho}  
~~~~~,~~~~~ \frac{1}{2m^*} = \frac{\partial \epsilon}{\partial\tau}
\label{skpot1}
\eeqa
which yields the single particle potential
\beqa
U_{\rm s.p.}(k_F,{\bf k}) = U_{\rm loc}(k_F) 
+ {\bf k}^2 \left(\frac{1}{2m^*}-\frac{1}{2M}\right) 
=  t_0 \varrho + (t_1 + t_2)\tau + {\bf k}^2 (t_1 + t_2)\varrho ~~~. 
\label{skpot2}
\eeqa
The energy density in non-relativistic colliding matter, 
i.e. for two separated spheres ($Mu >2 k_F$) with 
subsystem densities $\varrho (k_{F})$, total density 
 $\varrho_{12} (k_{F_{tot}})= 2\varrho (k_{F})$ and c.m. 
streaming velocities $\pm u$ has the form \cite{neise90}
\beqa
\epsilon_{12} &=& \frac{\tau_{12} }{2M} + \pi_{12} \nonumber\\
&=& 
\frac{\kappa}{(2\pi)^3}\int \frac{d^3 {\bf k}~{\bf k}^2}{2M} \Theta_{12}({\bf k})
+ \frac{1}{2}  \frac{\kappa^2}{(2\pi)^6} \int d^3 {\bf k}~ d^3 {\bf q} 
V({\bf k},{\bf q};{\bf k},{\bf q}) \Theta_{12}({\bf k}) \Theta_{12}({\bf q})
\nonumber \\
&=& 2 \frac{\tau (k_F) }{2M} 
+ \varrho_{12} \frac{M u^2}{2} + 
\frac{1}{2} t_0 \varrho^{2}_{12} + (t_1 + t_2) \left[ \varrho_{12} 2\tau (k_F) 
+  (\varrho_{12} Mu)^2\right]
\nonumber \\
&=& \epsilon |_{2\kappa} + u^2 \left[ \varrho_{12} \frac{M}{2} + 
 \varrho_{12}^2 M^2 (t_1 + t_2)\right]~~~.
\label{sk4}
\eeqa
The different contributions in Eq.(\ref{sk4}) are easy to interpret: 
Like in the relativistic case the energy density can be decomposed into 
a static part $\epsilon |_{2\kappa}$ which is determined by the 
doubled phase space volume, and a velocity dependent part. The latter 
contains the kinetic energy of the relative motion and the 
potential energy which originates from the integration of the interaction 
over the momentum spread of the two spheres, i.e. the 
interaction between the two currents. With (\ref{skpot1}) one 
obtains the local part of the single particle potential as 
\beqa
U_{\rm loc~12}
= \frac{\partial \epsilon_{12}}{\partial\varrho_{12}}  
= U_{\rm loc~12} (k_F) + U_{\rm loc~12}^\prime  (k_F,u) ~~.
\label{skpot3}
\eeqa
>From (\ref{sk4}) and (\ref{skpot3}) one finds that the velocity 
independent part of $U_{\rm loc}$ 
\beqa
U_{\rm loc~12}(k_F)
= \frac{\partial \epsilon |_{2\kappa}}{\partial\varrho_{12}}  
= 2 U_{\rm loc}(k_{F}) 
\label{skpot4}
\eeqa
equals two times the potential in nuclear matter at half the total 
density, i.e. it behaves like the scalar 
part of the relativistic mean field. In both cases the differences 
between the local potentials derived in NM and CNM originate 
from non-linear density 
dependence, (\ref{skpot1}), respectively the momentum integration  
over the interaction which creates this density dependence. 
In the relativistic case the $M^*/E^*$ weights (\ref{dens1},\ref{rs1}) 
play this role. Thus the reduced Fermi pressure affects not only 
the kinetic but also the internal potential energy of the subsystems. 
Although the Skyrme force (\ref{sk1}) will lead to unrealistic 
quantitative predictions for CNM at large velocities (it 
was especially designed for small relative momenta) 
it serves as an instructive example for a momentum dependent two-body 
interaction. The same qualitative arguments are valid when a more realistic 
interaction, e.g. the Brueckner G-matrix, is used in (\ref{sk4}). 
A stronger binding in colliding nuclear matter has e.g. been 
observed in non-relativistic G-matrix calculations for 
colliding matter \cite{gmat2}, but without giving an interpretation. 

One should be aware that this type of 
phase space effects is not included in standard transport 
calculations for heavy ion collisions, 
even when momentum dependent interactions are 
used. Phenomenological potentials 
are usually composed by a local, density dependent potential 
and a non-local momentum dependent part \cite{welke88}. For a nucleon with 
momentum ${\bf k}$ the single particle potential derived from the 
phase space distribution $f({\bf q}) \propto \Theta_{12}({\bf q})$ 
reads 
\beqa
U_{\rm s.p.} (k_{F_{\rm tot}},{\bf k} ) &=& U_{\rm loc}(k_{F_{\rm tot}} ) 
+  \frac{4}{(2\pi)^3} \int d^3q ~
V({\bf k},{\bf q};k_{F_{\rm tot}} ) \Theta_{12}({\bf q})
\nonumber\\
&=& U_{\rm loc}(k_{F_{\rm tot}} ) + U_{\rm nonloc} (k_{F_{\rm tot}},{\bf k})~~.
\label{pot}
\eeqa
In transport calculations $U_{\rm nonloc}$ arises from the integration 
of an effective two-body interaction over the actual 
momentum distribution $f$. Thus  $U_{\rm nonloc}$ accounts 
properly for the spread in momentum space between projectile and 
target nucleons as well as for the reduced Fermi momenta inside the 
subsystems. The same procedure should be applied to the local part 
of the potential. However, in standard transport calculations 
the density dependence of the mean field 
is parameterized into $U_{\rm loc}$ as a function of the 
total density. This density dependence is decoupled from the 
momentum space anisotropy and differs from  (\ref{skpot3}). 
Consequently, $U_{\rm loc}(k_{F_{\rm tot}})$ reflects the density 
dependence of equilibrated matter but should be calculated consistently 
also for anisotropic momentum space configurations. 
%%%%%%%%%%%%%%%%%%%%%%%%%%%%%%%%%%%%%%%%%%%%%%%%%%%%%%%%%%%%%%%%%%%%%%%%%%%%%
\section{Summary}
%%%%%%%%%%%%%%%%%%%%%%%%%%%%%%%%%%%%%%%%%%%%%%%%%%%%%%%%%%%%%%%%%%%%%%%%%%%%%
The early as well as high density phase of relativistic 
heavy ion reactions from SIS up to AGS energies are to large 
extent governed by phase space configurations far 
from local equilibrium. The phase space of the participant matter can 
be approximated by counter-streaming or colliding nuclear 
matter. In the present work we discussed implications for the nuclear 
equation of state which occurs in such non-equilibrium 
configurations. We restricted the discussion to the temperature zero case. 

The EOS in colliding matter is determined by two competing effects: 
The separation of phase space reduces the Fermi pressure in the 
subsystems compared to ground state matter. The reduction of 
the internal subsystem Fermi 
momenta affects also the potential part of the energy density and 
leads to an enlarged binding energy in the total system. For 
scalar quantities, e.g. the effective mass, the separation of 
phase space has is (in mean field approximation) formaly 
equivalent to the introduction of an additional degree of freedom. 
Using a more realistic interaction derived from the in-medium Dirac-Brueckner 
T-matrix this effect is present as well, but modified by the superimposed 
explicit momentum dependence of the interaction. Due to the repulsive 
character of the nuclear forces at high momenta, the relative motion 
of the two currents leads, on the other hand, to a strong additional 
repulsion in the system. The latter effect originates from the interaction 
between projectile and target nucleons and is taken into account by 
transport calculation when momentum dependent forces are used. The 
modification of the internal mean field of the subsystems by the 
separation in phase space is, however, not included in standard 
transport calculations. The magnitude of the discussed 
non-equilibrium effects is model 
dependent, however, the qualitative picture is model independent. 

In order to discuss the consequences for the equation of state in 
non-equilibrium, an effective EOS for colliding matter has been constructed 
where the energy of the relative motion has been subtracted. It was found 
as a model independent feature, that the separation of phase space leads 
to a considerable softening of the EOS in colliding matter. This effect is 
present over a wide range of densities and streaming velocities. 
We conclude that the discussed non-equilibrium effects should be taken into account 
when transport calculations for heavy ion collisions 
are performed. \\
\begin{acknowledgments}
We would like to  thank H.H. Wolter for many discussions.
\end{acknowledgments}
%%%%%%%%%%%%%%%%%%%%%%%%%%%%%%%%%%%%%%%%%%%%%%%%%%%%%%%%%%%%%%%%%%%%%%%%%
%                                                                       %
%   BEGIN OF BIBLIOGRAPHY                                               %
%                                                                       %
%%%%%%%%%%%%%%%%%%%%%%%%%%%%%%%%%%%%%%%%%%%%%%%%%%%%%%%%%%%%%%%%%%%%%%%%%

%%%%%%%%%%%%%%%%%%%%%%%%%%%%%%%%%%%%%%%%%%%%%%%%%%%%%%%%%%%%%%
\end{document}